\documentclass[12pt]{article}
\usepackage{amsfonts}
\usepackage{epic}
\usepackage{eepic}
\newtheorem{definition}{Definition}[section]

\newtheorem{proposition}[definition]{Proposition}

\newcommand{\finproof}{{\hfill \rule{5pt}{5pt}}}

          \def\cE{{\cal E}}          \def\cF{{\cal F}}
          \def\cH{{\cal H}}          
          \def\cK{{\cal K}}           
                    
                    \def\cR{{\cal R}} 
\def\cS{{\cal S}}                    \def\cU{{\cal U}}
          \def\cW{{\cal W}}          \def\cX{{\cal X}}

\def\CC{{\mathbb C}}

\def\ZZ{{\mathbb Z}}
\def\II{{\mathbb I}}

\newcommand{\si}{\sigma}
\newcommand{\ep}{\epsilon}
\newcommand{\pr}{\prime}
\newcommand{\nn}{\nonumber}

\begin{document}

\pagestyle{empty}

\begin{center} 
\textsf{\Large 
{Multi-leg integrable ladder models}} 
  
\vspace{36pt}
{\bf D.~Arnaudon\footnote{e-mail:{\sl arnaudon@lapp.in2p3.fr}}},
{\bf A.~Sedrakyan\footnote{e-mail:{\sl sedrak@lx2.yerphi.am,
Permanent address: Yerevan Physics Institute, Armenia}}},
{\bf T.~Sedrakyan\footnote{e-mail:{\sl tigrans@moon.yerphi.am,
Permanent address: Yerevan Physics Institute, Armenia}}},

\vspace{30pt}

\emph{Laboratoire d'Annecy-le-Vieux de Physique Th{\'e}orique LAPTH}
\\
\emph{CNRS, UMR 5108, associ{\'e}e {\`a} l'Universit{\'e} de Savoie}
\\
\emph{BP 110, F-74941 Annecy-le-Vieux Cedex, France}
\\

\vfill
{\bf Abstract}
\end{center}

We construct integrable spin chains with inhomogeneous 
periodic disposition of the
anisotropy parameter. The periodicity holds for both auxiliary (space)
and quantum (time) directions. The integrability of the model is based
on a set of coupled Yang--Baxter equations. 
This construction yields $P$-leg integrable ladder Hamiltonians.
We analyse the corresponding quantum group symmetry.

\vfill
\rightline{LAPTH-944/02}
\rightline{hep-th/0210087}
\rightline{October 2002}

\newpage
\pagestyle{plain}
\setcounter{page}{1}

\section{Introduction}

\indent

In an attempt to formulate the Chalker--Coddington model 
in the action formalism with coherent states \cite{CC}, one is led 
to the use of Manhattan Lattice $ML$ \cite{S, S1}. 
One has also to consider the $R$-matrices
of the $XX$ model with alternating shifts of the spectral parameter.
Therefore the idea of a $\ZZ_2$ staggering of the spaces
along the chain and time directions naturally rises up.

Initiating this fact in a series 
of articles \cite{APSS, AASSS,ST, ASSS} the authors have proposed
a technique for constructing integrable models where
the model parameter $\Delta$ (the anisotropy parameter of the
$XXZ$ and anisotropic $t-J$ models) have a staggered disposition 
of the sign along both chain and time directions. Due to the 
staggered shift of the spectral parameter also considered there, these
models have led to Hamiltonians formulated
on two leg zig-zag quasi-one dimensional chains.

The technique is based on an appropriate modification
of the Yang--Baxter equations ($YBE$) (\cite{Bax, FT}) for the
$R$-matrices which are the conditions for commutativity
of the transfer matrices at different values of the spectral
parameter. The transfer matrix is defined as the staggered product 
of $R$-matrices, which have staggered sign of the anisotropy parameter
$\Delta$ and staggered shift of the spectral parameter
of the model along the chain. The shift of spectral parameter
in a product of $R$-matrices was considered in 
earlier works of several authors \cite{SW, dV}, by use of which
they have analysed the ordinary $XXZ$ model in the infinite limit
of the spectral parameter. Later, in the article \cite{Z},
this modified transfer matrices has been used in order to construct
integrable models on the ladder.

Integrable models on ladders were also constructed in various
articles \cite{Wa, Ko, Muu, Al, For} but our construction essentially
differs from those because it contains an inhomogeneity
of the anisotropy parameter along the chain.

In the article \cite{ASSS} we have analysed the quantum group structure
behind of our construction and have shown that it is a tensor product
$\cW \otimes \cU_{q,i}(gl(2))$ of a Weyl algebra and a two parameter
($q$ and $i$) deformation of $\cU(gl(2))$.

In the present article we go further in this direction by presenting
the corresponding modification of our technique to consider and construct
$P$-leg integrable models, based on the $SL(2)$ group (the 
generalisation of the 2-leg $XXZ$ case).

In the Section 2 we present the basic definitions of our construction,
namely the $R$-matrices, the shifts of the spectral parameter, the
transfer matrix and the corresponding modified $YBE$ as the
conditions of commutativity of the transfer matrix for 
different   values of the spectral parameters. We also present
the solution of the coupled $YBE$ for the $XXZ$ case.

In the Section 3 we calculate the Hamiltonian of our model, showing
that it can be considered on the $P$-leg ladder. We first prove 
that our transfer matrix at zero value of the spectral parameter
is proportional to the identity operator (in the braid formalism).
The Hamiltonian, as usual, is defined as the linear term in the spectral
parameter expansion of the transfer matrix.

In Section 4 we analyse the quantum group structure behind of our 
Yang--Baxter equations. We show that the algebra defined by the RLL
relations \cite{FRT} 
derived from the set of coupled Yang--Baxter equations 
can be regarded as direct product of a Weyl
algebra and the $q$-deformed $\cU_q(sl(2))$.

\section{Basic Definitions and the Yang--Baxter equations}
\setcounter{equation}{0}

\indent

Let us now consider $\ZZ_P$ graded quantum $V_{j,\rho}(v)$ 
(with $j=1,...,L$ as a chain index) and 
auxiliary $V_{a,\si}(u)$ spaces, where $\rho, \si =0,1,...,P-1$ are
the grading indices. We consider
$R$-matrices which act on the direct product
of  spaces $V_{a,\si}(u)$ and $ V_{j,\rho}(v)$, $(\si,\rho =0,1,...,P-1)$,
mapping them on the intertwined direct product of 
$V_{a,\si+1}(u)$ and $ V_{j,\rho+1}(v)$ spaces (we impose 
periodic boundary conditions $P \equiv 0$)
\begin{equation}
  \label{R1}
  R_{aj,\si \rho}\left( u,v\right):\quad V_{a,\si}(u)\otimes 
  V_{j,\rho}(v)\rightarrow V_{j,\rho+1}(v)\otimes V_{a,\si+1}(u).  
\end{equation}
\begin{definition}
  We introduce   two transmutation operations $\iota_1$
  and $\iota_2$ with the property $\iota_1^P=\iota_2^P=id$ 
  for the quantum and auxiliary spaces
  correspondingly, and mark the operators $R_{aj,\si\rho}$ as 
  follows
  \begin{eqnarray}
    \label{R2}
    R_{aj,00}&\equiv& R_{aj},\qquad R_{aj,01}\equiv
    R_{aj}^{\iota_1}\qquad...\qquad R_{aj,0(P-1)}\equiv R_{aj}^{\iota_1^{P-1}}
    ,\nn\\
    R_{aj,10}&\equiv& R_{aj}^{\iota_2},\qquad R_{aj,11}\equiv R_{aj}^{\iota_1 
      \iota_2}\qquad...\qquad R_{aj,1(P-1)}\equiv R_{aj}^{\iota_2
      \iota_1^{P-1}},
    \nn\\
    \vdots\\
    R_{aj,(P-1)0}&\equiv& R_{aj}^{\iota_2^{P-1}},\qquad R_{aj,(P-1) 1}
    \equiv R_{aj}^{\iota_1 \iota_2^{P-1} }
    ...\qquad R_{aj,(P-1)(P-1)}\equiv R_{aj}^{\iota_1^{P-1} \iota_2^{P-1}}\nn
    .
  \end{eqnarray}
\end{definition}
The introduction of the $\ZZ_P$ grading of quantum spaces 
in time direction means, 
that we have now $P$ monodromy operators $T_{\rho}, \rho=0,1,...,P-1$,
which act on the space 
\begin{equation}
  V_{\rho}(u)=\prod_{j=1}^L V_{j,\rho}(u)
\end{equation}
by mapping it on $V_{\rho+1}(u)=\prod_{j=1}^P V_{j,\rho+1}(u)$
\begin{equation}
  \label{T}
  T_\rho(v-u) \qquad : V_\rho(u) \rightarrow V_{\rho+1}(u), \qquad \qquad 
  \rho=0,1,..P-1.
\end{equation}

It is clear now, that the monodromy operator of the model, which is defined
by translational invariance in $P$-steps in the time direction and  
determines the partition function, is the product of $P$ monodromy operators
\begin{equation}
  \label{TT}
  T(u) = \prod_{\si=0}^{P-1} T_\si(u).
\end{equation}

The $\ZZ_P$ grading of the auxiliary spaces 
along the chain direction besides the $\iota$ operations defined upper
contained also a shift of the spectral parameter. Let us define the
following 
shift operations on the spectral parameter $u$
\begin{equation}
  \label{SH}
  \bar{u} = \bar u^{(1)} =  \theta +\ep u,\quad \cdots, \quad 
  \bar{u}^{(k)} = \theta +\ep \bar{u}^{(k-1)}
  \qquad k=0,1,2,\cdots,P,
\end{equation}
where $\ep= e^{2\pi i/P}$ is the $P^{th}$ root of unity and $\theta$ is
an additional model parameter.
It is easy to prove that $\bar u^{(P)} = \bar{u}^{(0)}= u$.

By use of this shift operations we define
the $T_\si(u)$ monodromy matrices
according to the following
\begin{definition}
  We define the monodromy operators $T_{\si}(u), \si=0,1,2...,P-1$ 
  as a product
  of the $R_{aj,\si \rho}(u)$  matrices:
  \begin{eqnarray}
    \label{T1}
    T_0(u)=\prod_{j=1}^{L/P} (R_{a,Pj}(u)
    R_{a,Pj+1}^{\iota_2}(\bar u)\cdots 
    R_{a,Pj+P-1}^{\iota_2^{P-1}}(\bar u^{(P-1)})),\nn\\
    T_1(u)=\prod_{j=1}^{L/P} (R_{a,Pj}(\bar u)
    R_{a,Pj+1}^{\iota_1 \iota_2}(\bar u^{(2)})\cdots 
    R_{a,Pj+P-1}^{\iota_1 \iota_2^{P-1}}(\bar{u}^{(P)})),\nn\\
    \vdots \\
    T_p(u)=\prod_{j=1}^{L/P} (R_{a,Pj}(\bar u^{(p)})
    R_{a,Pj+1}^{\iota_1^p \iota_2}(\bar{u}^{(p+1)})\cdots 
    R_{a,Pj+P-1}^{\iota_1^p \iota_2^{P-1}}(\bar{u}^{(p+P-1)})),\nn\\
    p = 0,1...,P-1.\nn
  \end{eqnarray}
\end{definition}

We will consider this monodromy operators in this paper
and analyse their commutativity next.

As it is well known in Bethe Ansatz Technique \cite{Bax,FT}, a
sufficient 
condition for the commutativity of transfer matrices $\tau(u)=
Tr T(u)$ with different spectral parameters is the Yang--Baxter
equation ($YBE$). In order to have a commutativity of transfer matrices
(\ref{TT}) for different values of the spectral parameter we demand 
the commutativity of transfer
matrices (\ref{T1}). Then the standard so called railway arguments
yields the following set of $P$ equations
\begin{eqnarray}
  \label{eq:YBE1}
  R_{12}(\bar u-v) R_{13}^{\iota_1}(\bar u) R_{23}(v) &=&
  R_{23}^{\iota_1}(v) R_{13}(\bar u) \tilde{R}^{(1)}_{12}(\bar u-v)
  \nn\\[3mm]
  \tilde{R}^{(1)}_{12}(\bar u - v) R_{13}^{\iota_1 \iota_2}(\bar u^{(2)}) 
  R_{23}^{\iota_2}(\bar v) &=&
  R_{23}^{\iota_1 \iota_2}(\bar v) R_{13}^{\iota_2}(\bar u^{(2)}) 
\tilde{R}^{(2)}_{12}(\bar u^{(2)}-\bar v) \;,\nn\\
&\vdots&\\
\tilde{R}^{(P-1)}_{12}(\bar u^{(P)} - \bar{v}^{(P-1)}) 
R_{13}^{\iota_1 \iota_2^{P-1}}(\bar u^{(P)}) 
  R_{23}^{{\iota_2}^{P-1}}(\bar{v}^{(P-1)}) &=&
  R_{23}^{\iota_1 \iota_2^{P-1}}(\bar{v}^{(P-1)}) 
R_{13}^{{\iota_2}^{P-1}}(\bar u^{(P)}) 
R_{12}(\bar{u}-v).\nn
\end{eqnarray}

It can be seen easily that, defining $^{\iota_2}$ and
$\tilde{R}^{(p)}$ as 
\begin{eqnarray}
\label{iota}
R^{\iota_2}(u)&=&R^{\iota_1}(\ep^{-1} u),\nn\\
\tilde{R}^{(p)}(u-v)&=& R^{\iota_1^p}(u-v),\qquad p=1,...,P-1,
\end{eqnarray}
all the equations (\ref{eq:YBE1}) are compatible and reducing to the first
one.

Now, in order to solve this equation
for the $R(u)$ above, we  follow a procedure which is the inverse of
the Baxterisation (debaxterisation) \cite{Jones}.  
$\cR(u)$ can indeed be written
\begin{equation}
  \label{eq:deBaxterise}
  R_{12}(u) = \frac{1}{2i} \left(z R_{12} - z^{-1} R_{21}^{-1} \right)
\end{equation}
with $z=e^{iu}$ and the constant $R_{12}$ and $R_{21}^{-1}$ matrices are
spectral parameter independent. Then the Yang--Baxter equations 
(\ref{eq:YBE1})
for the spectral parameter dependent $R$-matrix $R(u)$ and
$R^{\iota_1}(u)$ are
equivalent to the following equations for the constant $R$-matrices
\begin{eqnarray}
  \label{eq:YBEconst1}
  R_{12} R_{13}^{\iota_1} R_{23} &\!=\!&
  R_{23}^{\iota_1} R_{13} R_{12}^{\iota_1} \\
  \label{eq:YBEconst2}
  R_{12}^{\iota_1} R_{13} R_{23}^{\iota_1} &\!=\!&
  R_{23} R_{13}^{\iota_1} R_{12} \\
  \label{eq:YBEconst3}
  R_{12} \left(R_{31}^{\iota_1}\right)^{-1} R_{23} 
  - \left(R_{21}\right)^{-1} R_{13}^{\iota_1} \left(R_{32}\right)^{-1} &\!=\!&
  R_{23}^{\iota_1} \left(R_{31}\right)^{-1} R_{12}^{\iota_1} 
  - \left(R_{32}^{\iota_1}\right)^{-1} R_{13}
  \left(R_{21}^{\iota_1}\right)^{-1}  \\
  \label{eq:YBEconst4}
  R_{12}^{\iota_1} \left(R_{31}\right)^{-1} R_{23}^{\iota_1} 
  - \left(R_{21}^{\iota_1}\right)^{-1} R_{13}
  \left(R_{32}^{\iota_1}\right)^{-1} &\!=\!& 
  R_{23} \left(R_{31}^{\iota_1}\right)^{-1} R_{12} 
  - \left(R_{32}\right)^{-1} R_{13}^{\iota_1} \left(R_{21}\right)^{-1}
  .~~~~~~~~~ 
\end{eqnarray}

If this modified YBE's have a solution, then one can formulate
new integrable models on the basis of existing ones. 
We will hereafter give solutions of these YBE's based on  $\cU_q(sl(2))$
$R$-matrices.

A solution of
(\ref{eq:YBEconst1})--(\ref{eq:YBEconst4}) is then
given by 
\begin{eqnarray}
  \label{eq:constR}
  R &=& 
  \left(\begin{array}{llll}
      q & 0 & 0 & 0 \\
      0 & 1 & 0  & 0 \\
      0 & q-q^{-1} & 1 & 0 \\
      0 & 0 & 0 & q
    \end{array}\right) \;,\\[4mm]
  R^{\iota_1^p} &=& 
  \label{eq:constRi1}
  \left(\begin{array}{llll}
      q & 0 & 0 & 0 \\
      0 & \ep^p & 0  & 0 \\
      0 & q-q^{-1} & \ep^{-p} & 0 \\
      0 & 0 & 0 & q
    \end{array}\right) \;,
\end{eqnarray}
where (\ref{eq:constR}) is the usual $R$-matrix of $\cU_q(gl(2))$.

The solution obtained in \cite{APSS} in connection with the 
staggered XXZ model for $P=2$, can then be generalised to a 
solution of (\ref{eq:YBE1}) given by
\begin{eqnarray}
  \label{eq:R}
  R(u) &=& 
  \left(\begin{array}{llll}
      \sin(\lambda +u) & 0 & 0 & 0 \\
      0 & \sin(u) & e^{-iu}\sin(\lambda)  & 0 \\
      0 & e^{iu}\sin(\lambda) & \sin(u) & 0 \\
      0 & 0 & 0 & \sin(\lambda +u)
    \end{array}\right) \;,\\[4mm]
  R^{\iota_1^p}(u) &=& 
  \left(\begin{array}{llll}
      \sin(\lambda +u) & 0 & 0 & 0 \\
      0 & \ep^p\cdot \sin(u) & e^{-iu}\sin(\lambda)  & 0 \\
      0 & e^{iu}\sin(\lambda) & \ep^{-p} \cdot \sin(u) & 0 \\
      0 & 0 & 0 & \sin(\lambda +u)
    \end{array}\right),
\end{eqnarray}
(Notice that we introduced here the off-diagonal factors $e^{iu}$ and
$e^{-iu}$  not present in \cite{APSS} to allow the
decomposition (\ref{eq:deBaxterise}). They are nothing more than a rescaling
of the states or a simple gauge transformation.)
\\
Following the technique developed in the article \cite{AKSM} one can
fermionise the $R$-matrix of the $XXZ$ model by using for the 
$V_{a,\si}$ and $V_{j,\rho}$ spaces the Fock space  
of the Fermi fields
$c_i$, $c_i^+$ with basis vectors $| 0 \rangle_i$ and 
$| 1 \rangle_i$, for which
\begin{equation}
  \label{X2}
  (X_i)_a^{a^{\pr}}=\left(\begin{array}{ll}
      1-n_i & c_i^+ \nn\\
      c_i & n_i
    \end{array}\right),
\end{equation}
is the  Hubbard operator.
Then the fermionic $R$-operator is defined by use of (\ref{eq:R})
\begin{eqnarray}
  \label{R3}
  R_{aj} &=&\left(-1\right) ^{p({a_1})p(j_2)}
  \left(R_{aj}\right)^{a_2j_2}_{{a_1}{j_1}}X_{a_2}^{a_1}
  X_{j_2}^{j_1}\nn\\
  &=& a(u)\left[-n_j n_k +(1-n_j)(1-n_k)\right]+
  b(u)\left[n_j(1- n_k) +(1-n_j) n_k\right]\nn\\
  &+&c(u)\left[c_j^+ c_k + c_k^+ c_j\right],  
\end{eqnarray}
where $a(u)=\sin(\lambda+u),\;\;\; b(u)=\sin(u),\;\;\;
c(u)=\sin(\lambda)$.
\\
The corresponding expressions for the
$R^{\iota_1^k}\;\;(k=1,\cdots,P-1)$ operators are 
\begin{eqnarray}
  \label{R31}
  R^{\iota_1^k}_{aj} 
  &=& a(u)\left[-n_j n_k +(1-n_j)(1-n_k)\right]+
  b(u)\left[\ep^{-k} n_j(1- n_k) +\ep^k (1-n_j) n_k\right]\nn\\
  &+&c(u)\left[c_j^+ c_k + c_k^+ c_j\right].  
\end{eqnarray}

\section{Hamiltonian for the  $P$-leg model}
\setcounter{equation}{0}

\indent

Usually in order to calculate the 
Hamiltonian in an homogeneous chain we should expand
the transfer
matrix around the point where it becomes identity operator (in the
braid formalism). But since we have different shifts of the spectral
parameter in our model it is impossible to find a value of the spectral
parameter
such that all the $R$-matrices become identity. 
Therefore we choose $u=0$ as an expansion point, where only some
of the $R$-matrices become identity, while the others contain scatterings.
This is the origin of the appearance of the next to nearest neighbour
interaction terms
($NNN$)in the Hamiltonian and of the formulation of the model on the P-leg
quasi-one dimensional chain.

We should calculate the transfer matrix at the point $u=0$
and the linear term of its expansion over $u$. 

\begin{proposition}
The  transfer matrix $\check{T}(0)$ in the braid formalism
is proportional to the identity operator
\begin{equation}
\label{T0}
\check{T}(0) \simeq \II
\end{equation}
\end{proposition}
\textbf{Proof:}
The proof is based on the use of YBE at the point $u=0$ and
graphically represented in Figure  \ref{fig:3} in the $P=3$ case.

\def\Rinso{$R(0)$}
\def\Rinsa{$R^{\iota^2}(\epsilon \theta)$}
\def\Rinsb{$R^{\iota}(-\epsilon^2 \theta)$}
\def\Rinsc{$R^{\iota^2}(\theta)$}
\def\Rinsd{$R^{\iota}(-\epsilon \theta)$}
\def\Rinse{$R^{\iota^2}(\epsilon^2 \theta)$}
\def\Rinsf{$R^{\iota}(-\theta)$}

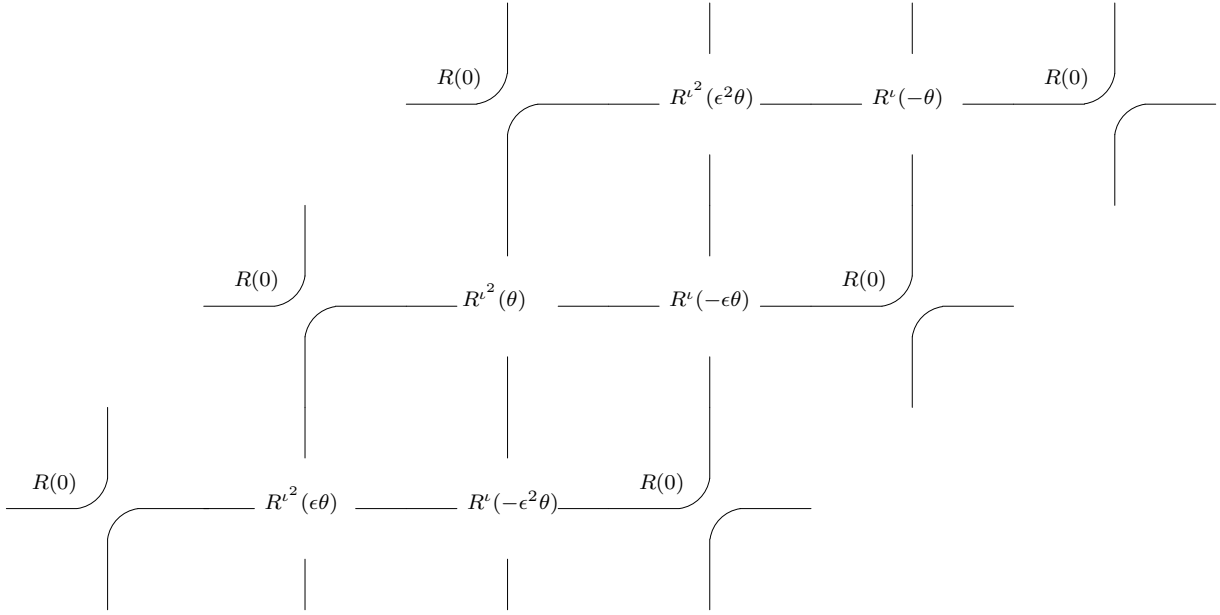
\begin{figure}[htbp]
  \centering
  
\setlength{\unitlength}{0.0015cm}
\begingroup\makeatletter\ifx\SetFigFont\undefined%
\gdef\SetFigFont#1#2#3#4#5{%
  \reset@font\fontsize{#1}{#2pt}%
  \fontfamily{#3}\fontseries{#4}\fontshape{#5}%
  \selectfont}%
\fi\endgroup%
{\renewcommand{\dashlinestretch}{30}
\begin{picture}(10779,5439)(0,-10)
\put(4152.000,4827.000){\arc{636.396}{0.1419}{1.4289}}
\put(4782.000,4197.000){\arc{636.396}{3.2835}{4.5705}}
\put(2352.000,3027.000){\arc{636.396}{0.1419}{1.4289}}
\put(2982.000,2397.000){\arc{636.396}{3.2835}{4.5705}}
\put(597.000,1227.000){\arc{636.396}{0.1419}{1.4289}}
\put(1227.000,597.000){\arc{636.396}{3.2835}{4.5705}}
\put(9552.000,4827.000){\arc{636.396}{0.1419}{1.4289}}
\put(10182.000,4197.000){\arc{636.396}{3.2835}{4.5705}}
\put(7752.000,3027.000){\arc{636.396}{0.1419}{1.4289}}
\put(8382.000,2397.000){\arc{636.396}{3.2835}{4.5705}}
\put(5952.000,1227.000){\arc{636.396}{0.1419}{1.4289}}
\put(6582.000,597.000){\arc{636.396}{3.2835}{4.5705}}
\path(2667,1812)(2667,1362)
\path(1767,912)(2217,912)
\path(3117,912)(3567,912)
\path(2667,462)(2667,12)
\path(4467,3612)(4467,3162)
\path(3567,2712)(4017,2712)
\path(4917,2712)(5367,2712)
\path(4467,2262)(4467,1812)
\path(4467,1812)(4467,1362)
\path(3567,912)(4017,912)
\path(4917,912)(5367,912)
\path(4467,462)(4467,12)
\path(6267,5412)(6267,4962)
\path(5367,4512)(5817,4512)
\path(6717,4512)(7167,4512)
\path(6267,4062)(6267,3612)
\path(8067,5412)(8067,4962)
\path(7167,4512)(7617,4512)
\path(8517,4512)(8967,4512)
\path(8067,4062)(8067,3612)
\path(6267,3612)(6267,3162)
\path(5367,2712)(5817,2712)
\path(6717,2712)(7167,2712)
\path(6267,2262)(6267,1812)
\path(4467,5412)(4467,4782)
\path(3567,4512)(4197,4512)
\path(4467,3612)(4467,4242)
\path(4737,4512)(5367,4512)
\path(2667,3612)(2667,2982)
\path(1767,2712)(2397,2712)
\path(2667,1812)(2667,2442)
\path(2937,2712)(3567,2712)
\path(912,1812)(912,1182)
\path(12,912)(642,912)
\path(912,12)(912,642)
\path(1182,912)(1812,912)
\path(9867,5412)(9867,4782)
\path(8967,4512)(9597,4512)
\path(9867,3612)(9867,4242)
\path(10137,4512)(10767,4512)
\path(8067,3612)(8067,2982)
\path(7167,2712)(7797,2712)
\path(8067,1812)(8067,2442)
\path(8337,2712)(8967,2712)
\path(6267,1812)(6267,1182)
\path(5367,912)(5997,912)
\path(6267,12)(6267,642)
\path(6537,912)(7167,912)
\put(4107,912){\makebox(0,0)[lb]{\smash{{{\SetFigFont{8}{10}{\rmdefault}{\mddefault}{\updefault}\Rinsb}}}}}
\put(4062,2712){\makebox(0,0)[lb]{\smash{{{\SetFigFont{8}{10}{\rmdefault}{\mddefault}{\updefault}\Rinsc}}}}}
\put(5907,2712){\makebox(0,0)[lb]{\smash{{{\SetFigFont{8}{10}{\rmdefault}{\mddefault}{\updefault}\Rinsd}}}}}
\put(5907,4512){\makebox(0,0)[lb]{\smash{{{\SetFigFont{8}{10}{\rmdefault}{\mddefault}{\updefault}\Rinse}}}}}
\put(7707,4512){\makebox(0,0)[lb]{\smash{{{\SetFigFont{8}{10}{\rmdefault}{\mddefault}{\updefault}\Rinsf}}}}}
\put(2307,912){\makebox(0,0)[lb]{\smash{{{\SetFigFont{8}{10}{\rmdefault}{\mddefault}{\updefault}\Rinsa}}}}}
\put(3837,4692){\makebox(0,0)[lb]{\smash{{{\SetFigFont{8}{10}{\rmdefault}{\mddefault}{\updefault}\Rinso}}}}}
\put(9237,4692){\makebox(0,0)[lb]{\smash{{{\SetFigFont{8}{10}{\rmdefault}{\mddefault}{\updefault}\Rinso}}}}}
\put(2037,2892){\makebox(0,0)[lb]{\smash{{{\SetFigFont{8}{10}{\rmdefault}{\mddefault}{\updefault}\Rinso}}}}}
\put(7437,2892){\makebox(0,0)[lb]{\smash{{{\SetFigFont{8}{10}{\rmdefault}{\mddefault}{\updefault}\Rinso}}}}}
\put(5637,1092){\makebox(0,0)[lb]{\smash{{{\SetFigFont{8}{10}{\rmdefault}{\mddefault}{\updefault}\Rinso}}}}}
\put(237,1092){\makebox(0,0)[lb]{\smash{{{\SetFigFont{8}{10}{\rmdefault}{\mddefault}{\updefault}\Rinso}}}}}
\end{picture}
}
  \caption{Transfer matrix at $u=0$ for $P=3$}
  \label{fig:3}
\end{figure}
 
\noindent
$T(u)$ is a product of $R$-matrices that can be written in several
ways, simply using the trivial commutations of matrices that have no
indices in common. On way to write $T(0)$ is
\begin{equation}
  \label{eq:T0a}
  T(0) = 
  \prod_{j=2}^{P} \; \prod_{i=1}^{j-1} 
  R_{ij}^{\iota^{P-j+i}}((\epsilon^{-j}-\epsilon^{-i})\theta')
  \;\; \prod_{j=1}^{P-1} \; \prod_{i=j+1}^{P} 
  R_{ij}^{\iota^{P-j+i}}((\epsilon^{-j}-\epsilon^{-i})\theta')
\end{equation}
where $\theta'=\theta/(1-\epsilon)$ and $\iota\equiv\iota_1$.
The first terms of this product are $R_{12}R_{13}R_{23}$ (with
convenient arguments and powers of 
$\iota$), for which we use the Yang--Baxter equation. 
$R_{12}$ meets then $R_{14}R_{34}$ and $R_{13}$ meets
$R_{14}R_{34}$. And so on, until all the $R_{1j}$ matrices are in the
middle (between the double products), so that
\begin{equation}
  \label{eq:T0b}
  T(0) = 
  \prod_{j=3}^{P} \; \prod_{i=2}^{j-1} 
  R_{ij}^{\iota^{P-j+i-1}}((\epsilon^{-j}-\epsilon^{-i})\theta')
  \prod_{j=2,..,k}^{\longleftarrow} R_{1j}^{\iota^{P+j-1}}
  \;\; \prod_{j=1}^{P-1} \; \prod_{i=j+1}^{P} 
  R_{ij}^{\iota^{P-j+i}}((\epsilon^{-j}-\epsilon^{-i})\theta')
\end{equation}
The $R_{i1}$ terms present in the last double product precisely
cancels the product 
$\prod\limits_{j=2,..,k}^{\longleftarrow} R_{1j}^{\iota^{P+j-1}}$ 
by unitarity property, so that 
\begin{equation}
  \label{eq:T0c}
  T(0) = 
  \prod_{j=3}^{P} \; \prod_{i=2}^{j-1} 
  R_{ij}^{\iota^{P-j+i-1}}((\epsilon^{-j}-\epsilon^{-i})\theta')
  \;\; \prod_{j=2}^{P-1} \; \prod_{i=j+1}^{P} 
  R_{ij}^{\iota^{P-j+i}}((\epsilon^{-j}-\epsilon^{-i})\theta')
\end{equation}
The powers of $\iota$ have changed in the remaining part. Following a
straightforward recursion, we get 
\begin{equation}
  \label{eq:T0d}
  T(0) = 
  \prod_{j=l+1}^{P} \; \prod_{i=l}^{j-1} 
  R_{ij}^{\iota^{P-j+i-l+1}}((\epsilon^{-j}-\epsilon^{-i})\theta')
  \;\; \prod_{j=l}^{P-1} \; \prod_{i=j+1}^{P} 
  R_{ij}^{\iota^{P-j+i}}((\epsilon^{-j}-\epsilon^{-i})\theta')
\end{equation}
and finally $T(0)=1$ (up to some translations).

\finproof

\paragraph{Hamiltonian for the 3-leg model}$ $
\\
We will restricts ourselves to $P = 3$ in the parts of this article 
concerned with the calculation of the Hamiltonian (the section
concerned by the underlying algebra deals with general $P$).
 
The transfer matrix $T(u)=T_0(u)T_1(u)T_2(u)$ for $P=3$ is defined by
formulae  (\ref{T1}) containing only three lines 
\begin{eqnarray}
  \label{T2}
  T_0(u)=\prod_{j=1}^{L/3} (R_{a,3j}(u)
  R_{a,3j+1}^{\iota_2}(\bar u) 
  R_{a,3j+2}^{\iota_2^{2}}(\bar u^{(2)})),\nn\\
  T_1(u)=\prod_{j=1}^{L/3} (R^{\iota_1}_{a,3j}(\bar u)
  R_{a,3j+1}^{\iota_1 \iota_2}(\bar u^{(2)}) 
  R_{a,3j+2}^{\iota_1 \iota_2^2}({u})),\\
  T_2(u)=\prod_{j=1}^{L/3} (R^{\iota_1^2}_{a,3j}(\bar u^{(2)})
  R_{a,3j+1}^{\iota_1^2 \iota_2}(u)
  R_{a,3j+2}^{\iota_1^2 \iota_2^{2}}(\bar{u})).\nn
\end{eqnarray}
For convenience let us use the graphical representations
of the $R$-matrices as it is defined in the article \cite{APSS}
and represented  in Figure  \ref{fig:1}. 

\medskip

\def\iPsii{$\psi_i$}
\def\iPsij{$\bar \psi_j$}
\def\iPsik{$\bar \psi_i$}
\def\iPsil{$\psi_j$}


\begin{figure}[htbp]
  \centering
  \setlength{\unitlength}{0.00083333in}
\begingroup\makeatletter\ifx\SetFigFont\undefined%
\gdef\SetFigFont#1#2#3#4#5{%
  \reset@font\fontsize{#1}{#2pt}%
  \fontfamily{#3}\fontseries{#4}\fontshape{#5}%
  \selectfont}%
\fi\endgroup%
{\renewcommand{\dashlinestretch}{30}
\begin{picture}(3684,1575)(0,-10)
\put(3300,1425){\makebox(0,0)[lb]{\smash{{{\SetFigFont{12}{14.4}{\rmdefault}{\mddefault}{\updefault}\iPsik}}}}}
\put(3300,225){\makebox(0,0)[lb]{\smash{{{\SetFigFont{12}{14.4}{\rmdefault}{\mddefault}{\updefault}\iPsil}}}}}
\put(225,900){\makebox(0,0)[lb]{\smash{{{\SetFigFont{12}{14.4}{\rmdefault}{\mddefault}{\updefault}$-b$}}}}}
\put(1800,1200){\makebox(0,0)[lb]{\smash{{{\SetFigFont{12}{14.4}{\rmdefault}{\mddefault}{\updefault}$a$}}}}}
\put(1800,0){\makebox(0,0)[lb]{\smash{{{\SetFigFont{12}{14.4}{\rmdefault}{\mddefault}{\updefault}$a$}}}}}
\put(3150,900){\makebox(0,0)[lb]{\smash{{{\SetFigFont{12}{14.4}{\rmdefault}{\mddefault}{\updefault}$b$}}}}}
\path(600,1425)(3000,1425)(3000,225)
        (600,225)(600,1425)
\path(600,1425)(1800,1425)
\path(1680.000,1395.000)(1800.000,1425.000)(1680.000,1455.000)
\path(3000,225)(3000,825)
\path(3030.000,705.000)(3000.000,825.000)(2970.000,705.000)
\path(3000,225)(1800,225)
\path(1920.000,255.000)(1800.000,225.000)(1920.000,195.000)
\path(600,1425)(600,825)
\path(570.000,945.000)(600.000,825.000)(630.000,945.000)
\put(0,1425){\makebox(0,0)[lb]{\smash{{{\SetFigFont{12}{14.4}{\rmdefault}{\mddefault}{\updefault}\iPsii}}}}}
\put(0,225){\makebox(0,0)[lb]{\smash{{{\SetFigFont{12}{14.4}{\rmdefault}{\mddefault}{\updefault}\iPsij}}}}}
\end{picture}
}
  \caption{$R_{ij}$}
  \label{fig:1}
\end{figure}

Later we will pass to the so called coherent state basis for the $R$-matrices
represented via Grassmann variables $\psi$, by which 
the corners of the square in the picture are labelled.
The transfer matrix $T(u)=T_0(u)T_1(u)T_2(u)$ defined by formulae 
(\ref{T1}) contains only three lines and can be drawn as in
Figure  \ref{fig:2}. 

\setlength{\unitlength}{0.0011cm}
\def\Rintaa{$R^{}(u)$}
\def\Rintab{$R^{\iota_2}(u)$}
\def\Rintac{$R^{\iota_2^2}(u)$}
\def\Rintba{$R^{\iota_1}(u)$}
\def\Rintbb{$R^{\iota_1\iota_2}(u)$}
\def\Rintbc{$R^{\iota_1\iota_2^2}(u)$}
\def\Rintca{$R^{\iota_1^2}(u)$}
\def\Rintcb{$R^{\iota_1^2\iota_2}(u)$}
\def\Rintcc{$R^{\iota_1^2\iota_2^2}(u)$}
\def\RintTa{$T_0$}
\def\RintTb{$T_1$}
\def\RintTc{$T_2$}

\begin{figure}
\label{fig:2}
\begingroup\makeatletter\ifx\SetFigFont\undefined%
\gdef\SetFigFont#1#2#3#4#5{%
  \reset@font\fontsize{#1}{#2pt}%
  \fontfamily{#3}\fontseries{#4}\fontshape{#5}%
  \selectfont}%
\fi\endgroup%
{\renewcommand{\dashlinestretch}{30}
\begin{picture}(13322,6090)(0,-10)
\path(11700,3510)(11250,3060)
\path(12600,3510)(13050,3060)
\path(13050,3060)(12600,2610)
\path(11250,3060)(11700,2610)
\path(11763.640,3616.066)(11700.000,3510.000)(11806.066,3573.640)
\path(11700,3510)(12150,3960)
\path(12150,3960)(12600,3510)
\path(12493.934,3573.640)(12600.000,3510.000)(12536.360,3616.066)
\path(12536.360,2503.934)(12600.000,2610.000)(12493.934,2546.360)
\path(12600,2610)(12150,2160)
\path(12150,2160)(11700,2610)
\path(11806.066,2546.360)(11700.000,2610.000)(11763.640,2503.934)
\path(11250,1260)(11700,1710)
\path(11636.360,1603.934)(11700.000,1710.000)(11593.934,1646.360)
\path(12706.066,1646.360)(12600.000,1710.000)(12663.640,1603.934)
\path(12600,1710)(13050,1260)
\path(13050,1260)(12600,810)
\path(12663.640,916.066)(12600.000,810.000)(12706.066,873.640)
\path(11593.934,873.640)(11700.000,810.000)(11636.360,916.066)
\path(11700,810)(11250,1260)
\path(11700,1710)(12150,2160)
\path(12150,2160)(12600,1710)
\path(12600,810)(12150,360)
\path(12150,360)(11700,810)
\path(11250,4860)(11700,5310)
\path(11636.360,5203.934)(11700.000,5310.000)(11593.934,5246.360)
\path(12706.066,5246.360)(12600.000,5310.000)(12663.640,5203.934)
\path(12600,5310)(13050,4860)
\path(13050,4860)(12600,4410)
\path(12663.640,4516.066)(12600.000,4410.000)(12706.066,4473.640)
\path(11593.934,4473.640)(11700.000,4410.000)(11636.360,4516.066)
\path(11700,4410)(11250,4860)
\path(11700,5310)(12150,5760)
\path(12600,4410)(12150,3960)
\path(12150,3960)(11700,4410)
\path(12150,5760)(12600,5310)
\path(4050,4860)(4500,5310)
\path(4436.360,5203.934)(4500.000,5310.000)(4393.934,5246.360)
\path(5506.066,5246.360)(5400.000,5310.000)(5463.640,5203.934)
\path(5400,5310)(5850,4860)
\path(5850,4860)(5400,4410)
\path(5463.640,4516.066)(5400.000,4410.000)(5506.066,4473.640)
\path(4393.934,4473.640)(4500.000,4410.000)(4436.360,4516.066)
\path(4500,4410)(4050,4860)
\path(4500,5310)(4950,5760)
\path(4950,5760)(5400,5310)
\path(5400,4410)(4950,3960)
\path(4950,3960)(4500,4410)
\path(900,3510)(450,3060)
\path(1800,3510)(2250,3060)
\path(2250,3060)(1800,2610)
\path(450,3060)(900,2610)
\path(963.640,3616.066)(900.000,3510.000)(1006.066,3573.640)
\path(900,3510)(1350,3960)
\path(1350,3960)(1800,3510)
\path(1693.934,3573.640)(1800.000,3510.000)(1736.360,3616.066)
\path(1736.360,2503.934)(1800.000,2610.000)(1693.934,2546.360)
\path(1800,2610)(1350,2160)
\path(1350,2160)(900,2610)
\path(1006.066,2546.360)(900.000,2610.000)(963.640,2503.934)
\path(450,1260)(900,1710)
\path(836.360,1603.934)(900.000,1710.000)(793.934,1646.360)
\path(1906.066,1646.360)(1800.000,1710.000)(1863.640,1603.934)
\path(1800,1710)(2250,1260)
\path(2250,1260)(1800,810)
\path(1863.640,916.066)(1800.000,810.000)(1906.066,873.640)
\path(793.934,873.640)(900.000,810.000)(836.360,916.066)
\path(900,810)(450,1260)
\path(900,1710)(1350,2160)
\path(1350,2160)(1800,1710)
\path(1800,810)(1350,360)
\path(1350,360)(900,810)
\path(2250,3060)(2700,3510)
\path(2636.360,3403.934)(2700.000,3510.000)(2593.934,3446.360)
\path(3706.066,3446.360)(3600.000,3510.000)(3663.640,3403.934)
\path(3600,3510)(4050,3060)
\path(4050,3060)(3600,2610)
\path(3663.640,2716.066)(3600.000,2610.000)(3706.066,2673.640)
\path(2593.934,2673.640)(2700.000,2610.000)(2636.360,2716.066)
\path(2700,2610)(2250,3060)
\path(2700,3510)(3150,3960)
\path(3150,3960)(3600,3510)
\path(3600,2610)(3150,2160)
\path(3150,2160)(2700,2610)
\path(2700,1710)(2250,1260)
\path(3600,1710)(4050,1260)
\path(4050,1260)(3600,810)
\path(2250,1260)(2700,810)
\path(2763.640,1816.066)(2700.000,1710.000)(2806.066,1773.640)
\path(2700,1710)(3150,2160)
\path(3150,2160)(3600,1710)
\path(3493.934,1773.640)(3600.000,1710.000)(3536.360,1816.066)
\path(3536.360,703.934)(3600.000,810.000)(3493.934,746.360)
\path(3600,810)(3150,360)
\path(3150,360)(2700,810)
\path(2806.066,746.360)(2700.000,810.000)(2763.640,703.934)
\path(4050,1260)(4500,1710)
\path(4436.360,1603.934)(4500.000,1710.000)(4393.934,1646.360)
\path(5506.066,1646.360)(5400.000,1710.000)(5463.640,1603.934)
\path(5400,1710)(5850,1260)
\path(5850,1260)(5400,810)
\path(5463.640,916.066)(5400.000,810.000)(5506.066,873.640)
\path(4393.934,873.640)(4500.000,810.000)(4436.360,916.066)
\path(4500,810)(4050,1260)
\path(4500,1710)(4950,2160)
\path(4950,2160)(5400,1710)
\path(5400,810)(4950,360)
\path(4950,360)(4500,810)
\path(4500,3510)(4050,3060)
\path(5400,3510)(5850,3060)
\path(5850,3060)(5400,2610)
\path(4050,3060)(4500,2610)
\path(4563.640,3616.066)(4500.000,3510.000)(4606.066,3573.640)
\path(4500,3510)(4950,3960)
\path(4950,3960)(5400,3510)
\path(5293.934,3573.640)(5400.000,3510.000)(5336.360,3616.066)
\path(5336.360,2503.934)(5400.000,2610.000)(5293.934,2546.360)
\path(5400,2610)(4950,2160)
\path(4950,2160)(4500,2610)
\path(4606.066,2546.360)(4500.000,2610.000)(4563.640,2503.934)
\path(2700,5310)(2250,4860)
\path(3600,5310)(4050,4860)
\path(4050,4860)(3600,4410)
\path(2250,4860)(2700,4410)
\path(2763.640,5416.066)(2700.000,5310.000)(2806.066,5373.640)
\path(2700,5310)(3150,5760)
\path(3150,5760)(3600,5310)
\path(3493.934,5373.640)(3600.000,5310.000)(3536.360,5416.066)
\path(3536.360,4303.934)(3600.000,4410.000)(3493.934,4346.360)
\path(3600,4410)(3150,3960)
\path(3150,3960)(2700,4410)
\path(2806.066,4346.360)(2700.000,4410.000)(2763.640,4303.934)
\path(450,4860)(900,5310)
\path(836.360,5203.934)(900.000,5310.000)(793.934,5246.360)
\path(1906.066,5246.360)(1800.000,5310.000)(1863.640,5203.934)
\path(1800,5310)(2250,4860)
\path(2250,4860)(1800,4410)
\path(1863.640,4516.066)(1800.000,4410.000)(1906.066,4473.640)
\path(793.934,4473.640)(900.000,4410.000)(836.360,4516.066)
\path(900,4410)(450,4860)
\path(900,5310)(1350,5760)
\path(1800,4410)(1350,3960)
\path(1350,3960)(900,4410)
\path(1350,5760)(1800,5310)
\path(9450,4860)(9900,5310)
\path(9836.360,5203.934)(9900.000,5310.000)(9793.934,5246.360)
\path(10906.066,5246.360)(10800.000,5310.000)(10863.640,5203.934)
\path(10800,5310)(11250,4860)
\path(11250,4860)(10800,4410)
\path(10863.640,4516.066)(10800.000,4410.000)(10906.066,4473.640)
\path(9793.934,4473.640)(9900.000,4410.000)(9836.360,4516.066)
\path(9900,4410)(9450,4860)
\path(9900,5310)(10350,5760)
\path(10350,5760)(10800,5310)
\path(10800,4410)(10350,3960)
\path(10350,3960)(9900,4410)
\path(6300,3510)(5850,3060)
\path(7200,3510)(7650,3060)
\path(7650,3060)(7200,2610)
\path(5850,3060)(6300,2610)
\path(6363.640,3616.066)(6300.000,3510.000)(6406.066,3573.640)
\path(6300,3510)(6750,3960)
\path(6750,3960)(7200,3510)
\path(7093.934,3573.640)(7200.000,3510.000)(7136.360,3616.066)
\path(7136.360,2503.934)(7200.000,2610.000)(7093.934,2546.360)
\path(7200,2610)(6750,2160)
\path(6750,2160)(6300,2610)
\path(6406.066,2546.360)(6300.000,2610.000)(6363.640,2503.934)
\path(5850,1260)(6300,1710)
\path(6236.360,1603.934)(6300.000,1710.000)(6193.934,1646.360)
\path(7306.066,1646.360)(7200.000,1710.000)(7263.640,1603.934)
\path(7200,1710)(7650,1260)
\path(7650,1260)(7200,810)
\path(7263.640,916.066)(7200.000,810.000)(7306.066,873.640)
\path(6193.934,873.640)(6300.000,810.000)(6236.360,916.066)
\path(6300,810)(5850,1260)
\path(6300,1710)(6750,2160)
\path(6750,2160)(7200,1710)
\path(7200,810)(6750,360)
\path(6750,360)(6300,810)
\path(7650,3060)(8100,3510)
\path(8036.360,3403.934)(8100.000,3510.000)(7993.934,3446.360)
\path(9106.066,3446.360)(9000.000,3510.000)(9063.640,3403.934)
\path(9000,3510)(9450,3060)
\path(9450,3060)(9000,2610)
\path(9063.640,2716.066)(9000.000,2610.000)(9106.066,2673.640)
\path(7993.934,2673.640)(8100.000,2610.000)(8036.360,2716.066)
\path(8100,2610)(7650,3060)
\path(8100,3510)(8550,3960)
\path(8550,3960)(9000,3510)
\path(9000,2610)(8550,2160)
\path(8550,2160)(8100,2610)
\path(8100,1710)(7650,1260)
\path(9000,1710)(9450,1260)
\path(9450,1260)(9000,810)
\path(7650,1260)(8100,810)
\path(8163.640,1816.066)(8100.000,1710.000)(8206.066,1773.640)
\path(8100,1710)(8550,2160)
\path(8550,2160)(9000,1710)
\path(8893.934,1773.640)(9000.000,1710.000)(8936.360,1816.066)
\path(8936.360,703.934)(9000.000,810.000)(8893.934,746.360)
\path(9000,810)(8550,360)
\path(8550,360)(8100,810)
\path(8206.066,746.360)(8100.000,810.000)(8163.640,703.934)
\path(9450,1260)(9900,1710)
\path(9836.360,1603.934)(9900.000,1710.000)(9793.934,1646.360)
\path(10906.066,1646.360)(10800.000,1710.000)(10863.640,1603.934)
\path(10800,1710)(11250,1260)
\path(11250,1260)(10800,810)
\path(10863.640,916.066)(10800.000,810.000)(10906.066,873.640)
\path(9793.934,873.640)(9900.000,810.000)(9836.360,916.066)
\path(9900,810)(9450,1260)
\path(9900,1710)(10350,2160)
\path(10350,2160)(10800,1710)
\path(10800,810)(10350,360)
\path(10350,360)(9900,810)
\path(9900,3510)(9450,3060)
\path(10800,3510)(11250,3060)
\path(11250,3060)(10800,2610)
\path(9450,3060)(9900,2610)
\path(9963.640,3616.066)(9900.000,3510.000)(10006.066,3573.640)
\path(9900,3510)(10350,3960)
\path(10350,3960)(10800,3510)
\path(10693.934,3573.640)(10800.000,3510.000)(10736.360,3616.066)
\path(10736.360,2503.934)(10800.000,2610.000)(10693.934,2546.360)
\path(10800,2610)(10350,2160)
\path(10350,2160)(9900,2610)
\path(10006.066,2546.360)(9900.000,2610.000)(9963.640,2503.934)
\path(8100,5310)(7650,4860)
\path(9000,5310)(9450,4860)
\path(9450,4860)(9000,4410)
\path(7650,4860)(8100,4410)
\path(8163.640,5416.066)(8100.000,5310.000)(8206.066,5373.640)
\path(8100,5310)(8550,5760)
\path(8550,5760)(9000,5310)
\path(8893.934,5373.640)(9000.000,5310.000)(8936.360,5416.066)
\path(8936.360,4303.934)(9000.000,4410.000)(8893.934,4346.360)
\path(9000,4410)(8550,3960)
\path(8550,3960)(8100,4410)
\path(8206.066,4346.360)(8100.000,4410.000)(8163.640,4303.934)
\path(5850,4860)(6300,5310)
\path(6236.360,5203.934)(6300.000,5310.000)(6193.934,5246.360)
\path(7306.066,5246.360)(7200.000,5310.000)(7263.640,5203.934)
\path(7200,5310)(7650,4860)
\path(7650,4860)(7200,4410)
\path(7263.640,4516.066)(7200.000,4410.000)(7306.066,4473.640)
\path(6193.934,4473.640)(6300.000,4410.000)(6236.360,4516.066)
\path(6300,4410)(5850,4860)
\path(6300,5310)(6750,5760)
\path(7200,4410)(6750,3960)
\path(6750,3960)(6300,4410)
\path(6750,5760)(7200,5310)
\put(1350,5940){\makebox(0,0)[lb]{\smash{{{\SetFigFont{8}{10}{\rmdefault}{\mddefault}{\updefault}-2}}}}}
\put(3150,5940){\makebox(0,0)[lb]{\smash{{{\SetFigFont{8}{10}{\rmdefault}{\mddefault}{\updefault}-1}}}}}
\put(4950,5940){\makebox(0,0)[lb]{\smash{{{\SetFigFont{8}{10}{\rmdefault}{\mddefault}{\updefault}0}}}}}
\put(6750,5940){\makebox(0,0)[lb]{\smash{{{\SetFigFont{8}{10}{\rmdefault}{\mddefault}{\updefault}1}}}}}
\put(8550,5940){\makebox(0,0)[lb]{\smash{{{\SetFigFont{8}{10}{\rmdefault}{\mddefault}{\updefault}2}}}}}
\put(10350,5940){\makebox(0,0)[lb]{\smash{{{\SetFigFont{8}{10}{\rmdefault}{\mddefault}{\updefault}3}}}}}
\put(12150,5940){\makebox(0,0)[lb]{\smash{{{\SetFigFont{8}{10}{\rmdefault}{\mddefault}{\updefault}4}}}}}
\put(1350,0){\makebox(0,0)[lb]{\smash{{{\SetFigFont{8}{10}{\rmdefault}{\mddefault}{\updefault}1}}}}}
\put(3150,0){\makebox(0,0)[lb]{\smash{{{\SetFigFont{8}{10}{\rmdefault}{\mddefault}{\updefault}2}}}}}
\put(4950,0){\makebox(0,0)[lb]{\smash{{{\SetFigFont{8}{10}{\rmdefault}{\mddefault}{\updefault}3}}}}}
\put(6750,0){\makebox(0,0)[lb]{\smash{{{\SetFigFont{8}{10}{\rmdefault}{\mddefault}{\updefault}4}}}}}
\put(8550,0){\makebox(0,0)[lb]{\smash{{{\SetFigFont{8}{10}{\rmdefault}{\mddefault}{\updefault}5}}}}}
\put(10350,0){\makebox(0,0)[lb]{\smash{{{\SetFigFont{8}{10}{\rmdefault}{\mddefault}{\updefault}6}}}}}
\put(12150,0){\makebox(0,0)[lb]{\smash{{{\SetFigFont{8}{10}{\rmdefault}{\mddefault}{\updefault}7}}}}}
\put(13230,4860){\makebox(0,0)[lb]{\smash{{{\SetFigFont{8}{10}{\rmdefault}{\mddefault}{\updefault}5}}}}}
\put(13230,3060){\makebox(0,0)[lb]{\smash{{{\SetFigFont{8}{10}{\rmdefault}{\mddefault}{\updefault}6}}}}}
\put(13230,1260){\makebox(0,0)[lb]{\smash{{{\SetFigFont{8}{10}{\rmdefault}{\mddefault}{\updefault}7}}}}}
\put(900,4860){\makebox(0,0)[lb]{\smash{{{\SetFigFont{8}{10}{\rmdefault}{\mddefault}{\updefault}\Rintab}}}}}
\put(2700,4860){\makebox(0,0)[lb]{\smash{{{\SetFigFont{8}{10}{\rmdefault}{\mddefault}{\updefault}\Rintac}}}}}
\put(4500,4860){\makebox(0,0)[lb]{\smash{{{\SetFigFont{8}{10}{\rmdefault}{\mddefault}{\updefault}\Rintaa}}}}}
\put(6300,4860){\makebox(0,0)[lb]{\smash{{{\SetFigFont{8}{10}{\rmdefault}{\mddefault}{\updefault}\Rintab}}}}}
\put(8100,4860){\makebox(0,0)[lb]{\smash{{{\SetFigFont{8}{10}{\rmdefault}{\mddefault}{\updefault}\Rintac}}}}}
\put(9900,4860){\makebox(0,0)[lb]{\smash{{{\SetFigFont{8}{10}{\rmdefault}{\mddefault}{\updefault}\Rintaa}}}}}
\put(11700,4860){\makebox(0,0)[lb]{\smash{{{\SetFigFont{8}{10}{\rmdefault}{\mddefault}{\updefault}\Rintab}}}}}
\put(11700,3060){\makebox(0,0)[lb]{\smash{{{\SetFigFont{8}{10}{\rmdefault}{\mddefault}{\updefault}\Rintbb}}}}}
\put(11700,1260){\makebox(0,0)[lb]{\smash{{{\SetFigFont{8}{10}{\rmdefault}{\mddefault}{\updefault}\Rintcb}}}}}
\put(9900,1260){\makebox(0,0)[lb]{\smash{{{\SetFigFont{8}{10}{\rmdefault}{\mddefault}{\updefault}\Rintca}}}}}
\put(9900,3060){\makebox(0,0)[lb]{\smash{{{\SetFigFont{8}{10}{\rmdefault}{\mddefault}{\updefault}\Rintba}}}}}
\put(8100,3060){\makebox(0,0)[lb]{\smash{{{\SetFigFont{8}{10}{\rmdefault}{\mddefault}{\updefault}\Rintbc}}}}}
\put(8100,1260){\makebox(0,0)[lb]{\smash{{{\SetFigFont{8}{10}{\rmdefault}{\mddefault}{\updefault}\Rintcc}}}}}
\put(6300,3060){\makebox(0,0)[lb]{\smash{{{\SetFigFont{8}{10}{\rmdefault}{\mddefault}{\updefault}\Rintbb}}}}}
\put(6300,1260){\makebox(0,0)[lb]{\smash{{{\SetFigFont{8}{10}{\rmdefault}{\mddefault}{\updefault}\Rintcb}}}}}
\put(4500,3060){\makebox(0,0)[lb]{\smash{{{\SetFigFont{8}{10}{\rmdefault}{\mddefault}{\updefault}\Rintba}}}}}
\put(4545,1260){\makebox(0,0)[lb]{\smash{{{\SetFigFont{8}{10}{\rmdefault}{\mddefault}{\updefault}\Rintca}}}}}
\put(2700,3060){\makebox(0,0)[lb]{\smash{{{\SetFigFont{8}{10}{\rmdefault}{\mddefault}{\updefault}\Rintbc}}}}}
\put(2700,1260){\makebox(0,0)[lb]{\smash{{{\SetFigFont{8}{10}{\rmdefault}{\mddefault}{\updefault}\Rintcc}}}}}
\put(900,3060){\makebox(0,0)[lb]{\smash{{{\SetFigFont{8}{10}{\rmdefault}{\mddefault}{\updefault}\Rintbb}}}}}
\put(855,1260){\makebox(0,0)[lb]{\smash{{{\SetFigFont{8}{10}{\rmdefault}{\mddefault}{\updefault}\Rintcb}}}}}
\put(0,4860){\makebox(0,0)[lb]{\smash{{{\SetFigFont{8}{10}{\rmdefault}{\mddefault}{\updefault}\RintTa}}}}}
\put(0,3060){\makebox(0,0)[lb]{\smash{{{\SetFigFont{8}{10}{\rmdefault}{\mddefault}{\updefault}\RintTb}}}}}
\put(0,1215){\makebox(0,0)[lb]{\smash{{{\SetFigFont{8}{10}{\rmdefault}{\mddefault}{\updefault}\RintTc}}}}}
\end{picture}
}

\caption{Triple row monodromy matrix}
\end{figure}
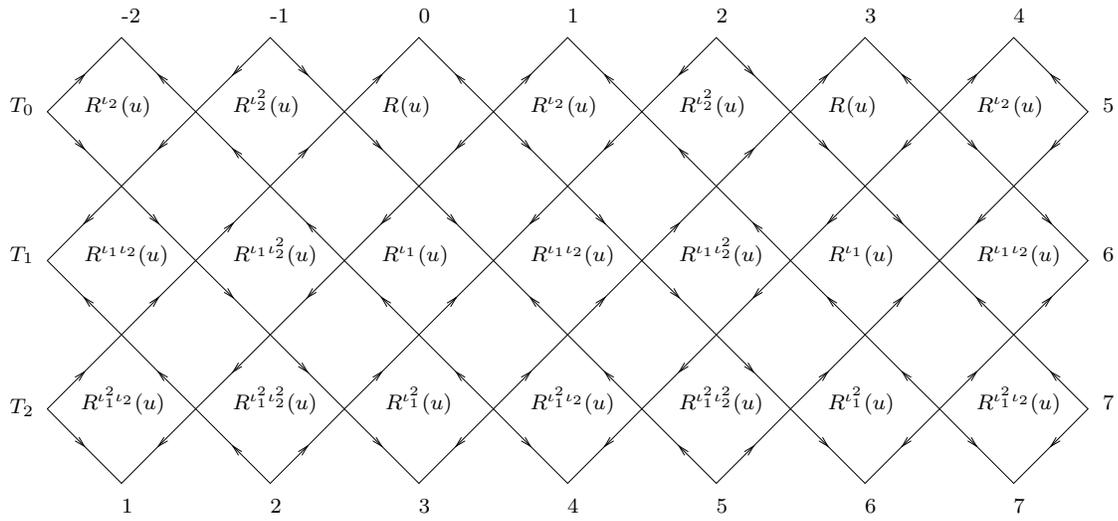

In order to calculate the Hamiltonian one should take the expansion
over the spectral parameter $u$
of the $R$-matrices in the expression (\ref{T2}) of the transfer
matrix, and extract the linear term. As one can see from Figure
\ref{fig:2}, we  
have nine  $R$-matrices forming a block, by translation
of which the whole transfer matrix can be reproduced. By use of
$T(0) \sim \II $ proved before one can obtain the following
expression for the contribution of one block to the Hamiltonian
\begin{eqnarray}
\label{Ham1}
T(u)&=&a_1a_2\cdots a_9(\II +u \cH_{01234}),\nn\\
\cH_{01234}&=& \cH^{(1)}_{01234}+\cH^{(2)}_{01234},\nn\\
\nn\\
\cH^{(1)}_{01234}&=&
H_{12}^{\iota_1^2\iota_2^2}
(\check{R}_{12}^{\iota_1^2\iota_2^2})^{-1}(\bar{0}^{(1)})
+(\check{R}_{23}^{\iota_2^2})^{-1}(\bar{0}^{(2)})
H_{23}^{\iota_2^2}\nn\\
&+&\check{R}_{12}^{\iota_1^2\iota_2^2}(\bar{0}^{(1)})
\check{R}_{23}^{\iota_1^2}(\bar{0}^{(2)})
H_{12}^{\iota_1}(\check{R}_{12}^{\iota_1})^{-1}(\bar{0}^{(1)})
(\check{R}_{23}^{\iota_2^2})^{-1}(\bar{0}^{(2)})
(\check{R}_{12}^{\iota_1^2\iota_2^2})^{-1}(\bar{0}^{(1)})\nn\\
&+&(\check{R}_{23}^{\iota_2^2})^{-1}(\bar{0}^{(2)})
(\check{R}_{12}^{\iota_2})^{-1}(\bar{0}^{(1)}) H_{12}^{\iota_2}
\check{R}_{23}^{\iota_2^2}(\bar{0}^{(2)})\nn\\
&+&\check{R}_{12}^{\iota_1^2\iota_2^2}(\bar{0}^{(1)}) H_{23}^{\iota_1^2}
(\check{R}_{23}^{\iota_1^2})^{-1}(\bar{0}^{(2)})
(\check{R}_{12}^{\iota_1^2\iota_2^2})^{-1}(\bar{0}^{(1)})\\ 
&+&(\check{R}_{23}^{\iota_2^2})^{-1}(\bar{0}^{(2)})
(\check{R}_{12}^{\iota_2})^{-1}(\bar{0}^{(1)})
(\check{R}_{23}^{\iota_1\iota_2})^{-1}(\bar{0}^{(2)})
H_{23}^{\iota_1\iota_2}\check{R}_{12}^{\iota_2}(\bar{0}^{(1)})
\check{R}_{23}^{\iota_2^2}(\bar{0}^{(2)}),\nn
\end{eqnarray}

\begin{eqnarray}
\label{Ham2}
\cH^{(2)}_{01234}&=&\check{R}_{12}^{\iota_1^2
  \iota_2^2}(\bar{0}^{(1)}) \check{R}_{23}^{\iota_1^2}(\bar{0}^{(2)})
H^{\iota_1^2 \iota_2}_{34} (\check{R}_{34}^{\iota_1^2 \iota_2})^{-1}(0) 
(\check{R}_{23}^{\iota_1^2})^{-1}(\bar{0}^{(2)})(\check{R}_{12}^{\iota_1^2
  \iota_2^2})^{-1}(\bar{0}^{(1)})\nn\\
&+& (\check{R}_{23}^{\iota_2^2})^{-1}(\bar{0}^{(2)})(\check{R}_{34}^{\iota_1
  \iota_2^2})^{-1}(0)
H^{\iota_1\iota_2^2}_{34}\check{R}_{23}^{\iota_2^2}(\bar{0}^{(2)})\nn\\
&+& (\check{R}_{12}^{\iota_1^2\iota_2^2})(\bar{0}^{(1)})
H^{\iota_1\iota_2^2}_{01}
(\check{R}_{01}^{\iota_1\iota_2^2})^{-1}(0)
(\check{R}_{12}^{\iota_1^2\iota_2^2})^{-1}(\bar{0}^{(1)})\\ 
&+& (\check{R}_{23}^{\iota_2^2})^{-1}(\bar{0}^{(2)})
(\check{R}_{12}^{\iota_2})^{-1}(\bar{0})
H_{34}\check{R}_{12}^{\iota_2})(\bar{0}
\check{R}_{23}^{\iota_2^2}(\bar{0}^{(2)})\nn\\
\nn
\end{eqnarray}
where the notation $\bar{0}^{(k)},\;k=1,2$ means $\bar{u}^{(k)}$
 at $u=0$ defined by the formula (\ref{SH}). In this expression
$H_{ij}^{\iota_1^p \iota_2^q}$ are the linear terms of the expansions
of the corresponding $R$-matrices and we will write down their
explicit expressions later.

The formulae (\ref{Ham1}) and (\ref{Ham2}) show that we
need the expressions of nine $R$-matrices, as well as their 
derivatives, at $u=0$. Instead of using in the future the
long upper scripts $\iota_1^k \iota_2^p$ it looks convenient to
 introduce the following numeration
for those nine $R$-matrices (see Figure  \ref{fig:2})
\begin{eqnarray}
\label{9R}
\begin{array}{ccc}
\check{R}^{\iota_2}(\bar{u}^{(1)})= R_1, & 
\check{R}^{\iota_2^2}(\bar{u}^{(2)})= R_4,&
\check{R}(u)=R_7 \\
\check{R}^{\iota_1}(\bar{u}^{(1)})= R_2,&
\check{R}^{\iota_1 \iota_2}(\bar{u}^{(2)})= R_5, & 
\check{R}^{\iota_1 \iota_2^2}(u)= R_8,\\
\check{R}^{\iota_1^2 \iota_2^2}(\bar{u}^{(1)})= R_3,&
\check{R}^{\iota_1^2}(\bar{u}^{(2)})= R_6, & 
\check{R}^{\iota_1^2 \iota_2}(u)= R_9,
\end{array}
\end{eqnarray}
with the corresponding numerations for the $a(0), \;\; b(0)$
parameters(the parameter $c$
is the same for all of $R$-matrices and equal to $c=\cos\lambda$.
\begin{eqnarray}
\label{9ab}
\begin{array}{lll}
a_1=\sin(\ep^2 \theta+\lambda)&a_2=\sin(\theta+\lambda)&
a_3=\sin(\ep \theta+\lambda),\\
\vspace{0.5cm}
b_1=\sin(\ep^2 \theta)&b_2=\sin(\theta)&
b_3=\sin(\ep \theta),\\
a_4=\sin(- \theta+\lambda)&a_5=\sin(-\ep \theta+\lambda)&
a_6=\sin(-\ep^2 \theta+\lambda),\\
\vspace{0.5cm}
b_4=\sin(-\theta)&b_5=\sin(-\ep \theta)&
b_6=\sin(-\ep^2 \theta),\\
a_7=\sin(\lambda)&a_8=\sin(\lambda)&
a_9=\sin(\lambda),\\
b_7=0 & b_8=0&
b_9=0\\
\end{array}
\end{eqnarray}
and for their derivatives $a^{\pr}(0), \;\; b^{\pr}(0)$ 
\begin{eqnarray}
\label{9abp}
\begin{array}{lll}
a^{\pr}_1=\cos(\ep^2 \theta+\lambda)&a^{\pr}_2=\ep \cos(\theta+\lambda)&
a^{\pr}_3=\ep^2 \cos(\ep \theta+\lambda),\\
\vspace{0.5cm}
b^{\pr}_1=\cos(\ep^2 \theta)&b^{\pr}_2=\ep \cos(\theta)&
b^{\pr}_3=\ep^2 \cos(\ep \theta),\\
a^{\pr}_4=\cos(- \theta+\lambda)&a^{\pr}_5=\ep \cos(-\ep \theta+\lambda)&
a^{\pr}_6=\ep^2 \cos(-\ep^2 \theta+\lambda),\\
\vspace{0.5cm}
b^{\pr}_4=\cos(-\theta)&b^{\pr}_5=\ep \cos(-\ep \theta)&
b^{\pr}_6=\ep^2 \cos(-\ep^2 \theta).
\end{array}
\end{eqnarray}

\begin{figure}[htbp]
  \centering
\setlength{\unitlength}{0.0009cm}
\begingroup\makeatletter\ifx\SetFigFont\undefined%
\gdef\SetFigFont#1#2#3#4#5{%
  \reset@font\fontsize{#1}{#2pt}%
  \fontfamily{#3}\fontseries{#4}\fontshape{#5}%
  \selectfont}%
\fi\endgroup%
{\renewcommand{\dashlinestretch}{30}
\begin{picture}(13974,3639)(0,-10)
\put(169.500,2554.500){\arc{318.198}{2.9997}{4.8543}}
\put(5254.500,2869.500){\arc{318.198}{6.1413}{7.9959}}
\put(9619.500,2554.500){\arc{318.198}{2.9997}{4.8543}}
\put(12904.500,2869.500){\arc{318.198}{6.1413}{7.9959}}
\put(1069.500,1654.500){\arc{318.198}{2.9997}{4.8543}}
\put(3454.500,1969.500){\arc{318.198}{6.1413}{7.9959}}
\put(4354.500,1069.500){\arc{318.198}{6.1413}{7.9959}}
\put(1969.500,754.500){\arc{318.198}{2.9997}{4.8543}}
\put(10519.500,1654.500){\arc{318.198}{2.9997}{4.8543}}
\put(12004.500,1969.500){\arc{318.198}{6.1413}{7.9959}}
\put(13804.500,1069.500){\arc{318.198}{6.1413}{7.9959}}
\put(12319.500,754.500){\arc{318.198}{2.9997}{4.8543}}
\path(2712,12)(2712,822)
\path(2712,1002)(2712,1722)
\path(2712,1902)(2712,2622)
\path(192,2712)(2622,2712)
\path(2712,2802)(2712,3612)
\path(2622,2802)(2802,2802)(2802,2622)
        (2622,2622)(2622,2802)
\path(2802,2712)(5277,2712)
\path(912,12)(912,1632)
\path(1812,12)(1812,732)
\path(12,2532)(12,12)
\path(3612,2802)(3612,3612)
\path(4512,2802)(4512,3612)
\path(5412,2892)(5412,3612)
\path(3612,1992)(3612,2622)
\path(4512,1092)(4512,2622)
\path(12162,12)(12162,732)
\path(9642,2712)(11172,2712)
\path(11352,2712)(12882,2712)
\path(10362,12)(10362,1632)
\path(11262,12)(11262,1722)
\path(11172,2802)(11352,2802)(11352,2622)
        (11172,2622)(11172,2802)
\path(11262,2802)(11262,3612)
\path(11262,1902)(11262,2622)
\path(9462,2532)(9462,12)
\path(12162,2802)(12162,3612)
\path(12162,1992)(12162,2622)
\path(13062,2892)(13062,3612)
\path(13962,1092)(13962,3612)
\path(1092,1812)(3432,1812)
\path(1992,912)(4332,912)
\path(10542,1812)(11982,1812)
\path(12342,912)(13782,912)
\end{picture}
}
  \caption{Some terms of the Hamiltonian}
  \label{fig:4}
\end{figure}
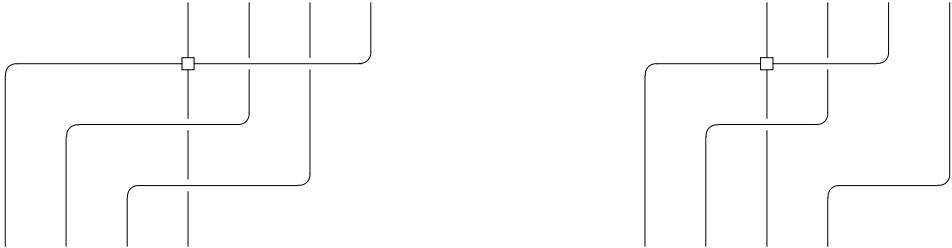

The Hamiltonian (\ref{Ham1}-\ref{Ham2}) can have a following simple
interpretation. We are taking the logarithmic derivatives of
the nine $R$-matrices of the constituent block, but instead of
ordinary
sum we should take the braided sums of this terms. The meaning is
clear from Figure  \ref{fig:4}, where the square box represents
$H$ and where crossings of lines represent $R$-matrices.  

First one should make the
permutations of neighbouring nodes 
with the corresponding $R^{\iota_1^k \iota_2^p}(0)$ matrix and after
the interaction via the local Hamiltonian  permute them back to their 
original order with the inverse matrices $(R^{\iota_1^k \iota_2^p})^{-1}(0)$.
This interpretation is interesting from the mathematical point
of view and would need further investigation.
Nevertheless it looks necessary to write down explicitly the
expression for the Hamiltonian in terms of fermionic
creation-annihilation operators.  

In order to make the calculations it appeared
to be convenient to pass from ordinary matrix (\ref{eq:R}) or operator
(\ref{R3}-\ref{R31}) expressions of the $R$-matrix to the so called
coherent state basis representation. 
Let us introduce the fermionic coherent states according to
articles \cite{F} and express the $R$-matrix in these terms
as it is done in \cite{S, S1}.
\begin{eqnarray}
  \label{CS1}
  |\psi_{2j}\rangle = e^{\psi_{2j}c^+_{2j}}|0\rangle ,\qquad 
  \langle\bar{\psi}_{2j}| = \langle0| e^{c_{2j}\bar{\psi}_{2j}}
\end{eqnarray}
for the even sites of the chain and
\begin{eqnarray}
  \label{CS2}
  |\bar{\psi}_{2j+1}\rangle = (c^+_{2j+1}-\bar{\psi}_{2j+1})
  |0\rangle ,\qquad
  \langle\psi_{2j+1}| = \langle 0|(c_{2j+1}+\psi_{2j+1})
\end{eqnarray}
for the odd sites. 

These states are designed as an eigenstates of
creation-annihilation operators of fermions $c_j^+,\;\; c_j $
with eigenvalues $\psi_j$ and $\bar\psi_j$
\begin{eqnarray}
  \label{CPR}
  c_{2j}\mid \psi_{2j}\rangle =- \psi_{2j}\mid \psi_{2j}\rangle &,&
  \langle\bar{\psi}_{2j} \mid c^+_{2j} = 
  -\langle\bar{\psi}_{2j} \mid \bar{\psi}_{2j},\\
  c^+_{2j+1}\mid \bar{\psi}_{2j+1}\rangle =
  \bar{\psi}_{2j+1}\mid \bar{\psi}_{2j+1}\rangle &,&
  \langle\psi_{2j+1}\mid c_{2j+1} =- \langle\psi_{2j+1}\mid \psi_{2j+1}.\nn
\end{eqnarray}
It is easy to calculate the scalar product of this states
\begin{eqnarray}
  \label{PP1}
  \langle \bar{\psi}_{2j}\mid \psi_{2j}\rangle =
  e^{\bar{\psi}_{2j} \psi_{2j}},\nn\\
  \label{TpR}
  \langle\psi_{2j+1} \mid \bar{\psi}_{2j+1}\rangle =
  e^{\bar{\psi}_{2j+1} \psi_{2j+1}}
\end{eqnarray}
and find the completeness relations  
\begin{eqnarray}
  \label{PP2}
  \int d\bar{\psi}_{2j}d\psi_{2j}\mid \psi_{2j}
  \rangle \langle\bar{\psi}_{2j}\mid e^{\psi_{2j}\bar{\psi}_{2j}}&=& 1,
  \nn\\
  \label{CRE}
  \int d\bar{\psi}_{2j+1}d\psi_{2j+1}\mid \bar{\psi}_{2j+1}\rangle 
  \langle\psi_{2j+1}\mid e^{\psi_{2j+1}\bar{\psi}_{2j+1}}&=& 1.
\end{eqnarray}
Now let us pass to the coherent basis 
in the spaces of states $\prod_j V_{j,\si}$ 
of the chain and calculate the matrix elements
of the $R_{2j,2j\pm1}$-operators between the initial
$\mid \psi_{2j}\rangle ,\;\; \mid \bar{\psi}_{2j\pm1}\rangle$
and final $\langle\bar\psi_{2j}^{\pr}\mid,
\;\;\langle\psi_{2j\pm1}^{\pr}\mid$
states. Using the properties of coherent states it is easy to find
from the formula (\ref{R3}), that 
\begin{eqnarray}
  \label{Rpsi}
  (R)
  _{\psi_{2j},\bar{\psi}_{2j\pm1}}^{\bar{\psi}^{\pr}_{2j},\psi^{\pr}_{2j\pm1}}
  (u)&=&
  \langle\psi^{\pr}_{2j\pm1},\bar{\psi}^{\pr}_{2j}\mid(\check{R}_k)_{2j,2j\pm1}
  \mid\psi_{2j},\bar{\psi}_{2j\pm1}\rangle \nn\\
  &=&e^{\left[a(u)\bar{\psi}^{\pr}_{2j}\psi_{2j}
      +a(u)\bar{\psi}_{2j\pm1}\psi^{\pr}_{2j\pm1}
      -b(u)\bar{\psi}_{2j\pm1}\psi_{2j} +
      b(u)\bar{\psi}^{\pr}_{2j}\psi^{\pr}_{2j\pm1}
      - \delta\bar{\psi}^{\pr}_{2j}\psi_{2j}
      \bar{\psi}_{2j\pm1}\psi^{\pr}_{2j\pm1} \right]},\nn\\
  && \qquad \qquad \qquad \qquad \qquad \qquad \qquad k=1,2...,9.
\end{eqnarray}
where $\delta =2 a(u) b(u) \Delta$ and $\Delta=\cos(\lambda)$ is the
models anisotropy parameter.

By use of these coherent states we can replace all the ordinary
matrix multiplications in the expressions
(\ref{Ham1}-\ref{Ham2}) of the Hamiltonian 
by the Grassmann variable integration
with the measure
\begin{equation}
\label{measure}
\int\cdots e^{\psi_i \bar{\psi}_i} d\bar{\psi}_i d\psi_i
\end{equation}
as it is dictated by the formulae (\ref{PP2}). Then the calculation of
integrals
over Grassmann variables simply reduces to the all possible Wilson--Polyakov's 
contour integral countings defined on a graph connected to the terms
in the expression (\ref{Ham1}-\ref{Ham1}) by use of
identification of the Figure  \ref{fig:1} with the $R$-matrix.

The expression we have obtained for the $\cH^{(1)}_{01234}$
is the following
\begin{eqnarray}
\label{cH1}
\cH^{(1)}_{01234}&=&
c_2^+ c_1(t_{21}+f_{21} n_3) + c_1^+ c_2(t_{12}+f_{12}n_3)
+c_2^+ c_3(t_{23}+f_{23} n_1) + c_3^+ c_2(t_{32}+f_{32}n_1)\nn\\
&+&c_3^+ c_1(t_{31}+f_{31} n_2)+c_1^+ c_3(t_{13}+f_{13}n_2)+d_{11}n_1+
d_{22}n_2 +d_{33}n_3-d_{12}n_1n_2\nn\\
&+&d_{13}n_1n_3-d_{23}n_2n_3
+d_{123}n_1n_2n_3,
\end{eqnarray}
where the coefficients $t_{ij},f_{ij},d_{ij},\;i,j=1,2,3$ and
$d_{123}$
are written down in the Appendix.

The expression for the second part of Hamiltonian $\cH^{(2)}_{01234}$
which consist of  derivatives of the $R_7(u),R_8(u),R_9(u)$
terms in the transfer matrix can be summarised as follows.
\begin{eqnarray}
\label{H3}
\cH^{(2)}_{01234}&=&H_{7,0}^0\cdot\cS_{1,1}^0\cdot\cS_{4,2}^0+
H_{7,0}^{\mu}\cdot(\cS_{1,1})_{\mu}^{nu}\cdot(\cS_{4,2})_{\nu}^{\si}
\cdot X_{3,\si}\nn\\
&+&\delta_7^{\pr}(\cS_{4,0}^T)^0\cdot\bar{\cS}_{3,1}^0+
X_{-1}^{\si}\cdot(\cS_{4,0}^T)_{\si}^{\mu}\cdot(H_{8})_{\mu}^{\nu}
\cdot(\bar{\cS}_{3,1})_{\nu}^{\rho}
\cdot X_{2,\rho}\nn\\
&+&(\bar{\cS}_{3,2}^T)^0 \cdot (\bar{\cS}_{6,3}^T)^0 \cdot H_{9,4}^0+
X_{1}^{\si}\cdot(\bar{\cS}_{3,2}^T)_{\si}^{\mu}\cdot
(\bar{\cS}_{6,3}^T)_{\mu}^{\nu}
\cdot (H_{9,4})_\nu,
\end{eqnarray}
where vector indices $\mu, \nu, \si = +, -, 3$ and can be taken up
and down by the usual metric tensor 
$g_{\mu \nu}=\left(\begin{array}{ccc}
0&2&0\\
2&0&0\\
0&0&1
\end{array}
\right)$, 
upper script $T$ means transposition. 

In the expression
(\ref{H3}) other notations are defined as
\begin{eqnarray}
\label{H4}
H_{r,i}^0&=&a_r^{\pr}-(\delta_r^{\pr}-a_r^{\pr})c_i^+c_i,\nn\\
H_{r,i}^{\mu}&=&(b_r^{\pr} c_i^+ , b_r^{\pr} c_i^, 
a_r^{\pr}-(\delta_r^{\pr}+a_r^{\pr})c_i^+c_i),\nn\\
(H_r)_{\mu}^{\nu}&=&
\left(\begin{array}{ccc}
2 b_r^{\pr}&0&0\\
0&-2b_r^{\pr}&0\\
0&0&\delta_r^{\pr}
\end{array}\right),\nn\\
X_{i}^{\mu}&=&(c_i,\;c_i^+,\; 1-2 n_i ) \qquad \qquad r=7,8,9 
\end{eqnarray}
and
\begin{eqnarray}
\label{cS}
\cS_{r,i}^0&=& 2 a_r \bar{a}_r e^{c_i^+c_i},\qquad \qquad r=7,8,9\nn\\
(\cS_{r,i})_{\mu}^{\nu}&=&
\left(\begin{array}{ccc}
-2 b_r \bar{a}_r e^{\frac{a_r}{\bar{a}_r}
  c_i^+c_i}&0&\sqrt{2}(\bar{a}_r-a_r)c_i\\
0&-2 b_r \bar{a}_r e^{\frac{\bar{a}_r}{a_r}
 c_i^+c_i}&\sqrt{2}(\bar{a}_r-a_r)c_i^+\\
-2 \sqrt{2} b_r c_i^+ & -2 \sqrt{2} b_r c_i & 2 b_r^2 e^{c_i^+c_i}
\end{array}\right)
\end{eqnarray}
where subscript $i$ enumerates the lattice site. In the formulae
(\ref{cS}) $\bar{a}_1=a_6,\;\bar{a}_2=a_4,\;\bar{a}_3=a_5,
\bar{a}_4=a_2,\;\bar{a}_5=a_3,\;\bar{a}_6=a_1$.

\section{Algebra}
\setcounter{equation}{0}

\indent

The underlying algebra of this model is defined by generators
$L^{\iota^p}$ and the following set
of RLL relations derived from the set of coupled Yang--Baxter
equations following \cite{FRT} 
\\
\begin{eqnarray}
  \label{eq:algebra}
  &&
  R_{12}^{\iota^p} {L^\pm_1}^{\iota^{p+1}} {L^\pm_2}^{\iota^p}
  =
  {L^\pm_2}^{\iota^{p+1}}  {L^\pm_1}^{\iota^{p}} R_{12}^{\iota^{p+1}}
  \\
  &&
  R_{12}^{\iota^p} {L^+_1}^{\iota^{p+1}} {L^-_2}^{\iota^p}
  =
  {L^-_2}^{\iota^{p+1}}  {L^+_1}^{\iota^{p}} R_{12}^{\iota^{p+1}}
\end{eqnarray}
(with again $\iota\equiv\iota_1$). 
Writing
\begin{equation}
  \label{eq:Lmatrix}
  {L^+}^{\iota^p} = \left(
    \begin{array}{cc}
      K_{+1^{\iota^p}} & 0 \\
      E^{\iota^p} & K_{+2}^{\iota^p} 
    \end{array}
  \right),
  \qquad
  {L^-}^{\iota^p} = \left(
    \begin{array}{cc}
      K_{-1}^{\iota^p} & ^{\iota^p}F \\
      0 & K_{-2}^{\iota^p} 
    \end{array}
  \right)
\end{equation}
these relations become
\begin{equation}
  \label{eq:KK}
  \begin{array}{l}
    K_{+1}^{\iota_1^{p+1}} K_{-1}^{\iota_1^p} 
    = K_{-1}^{\iota_1^{p+1}} K_{+1}^{\iota_1^p}
    \\[2mm]
    K_{+2}^{\iota_1^{p+1}} K_{-2}^{\iota_1^p}
    = K_{-2}^{\iota_1^{p+1}} K_{+2}^{\iota_1^p} 
    \\[2mm]
    K_{+1}^{\iota_1^{p+1}} K_{+2}^{\iota_1^p} 
    = \epsilon ~ K_{+2}^{\iota_1^{p+1}} K_{+1}^{\iota_1^p} 
    \\[2mm]
    K_{-1}^{\iota_1^{p+1}} K_{-2}^{\iota_1^p} 
    = \epsilon ~ K_{-2}^{\iota_1^{p+1}} K_{-1}^{\iota_1^p} 
    \\[2mm]
    K_{+1}^{\iota_1^{p+1}} K_{-2}^{\iota_1^p} 
    = \epsilon ~ K_{-2}^{\iota_1^{p+1}} K_{+1}^{\iota_1^p} 
    \\[2mm]
    K_{+2}^{\iota_1^{p+1}} K_{-1}^{\iota_1^p} 
    = \epsilon^{-1} ~ K_{-1}^{\iota_1^{p+1}} K_{+2}^{\iota_1^p} 
  \end{array}
\end{equation}
\begin{equation}
  \label{eq:KE}
  \begin{array}{l}
    K_{+1}^{\iota_1^{p+1}} E^{\iota_1^p} 
    = \epsilon^{-p} q E^{\iota_1^{p+1}} K_{+1}^{\iota_1^p} 
    \\[2mm]
    K_{-1}^{\iota_1^{p+1}} E^{\iota_1^p} 
    = \epsilon^{-p} q^{-1} E^{\iota_1^{p+1}} K_{-1}^{\iota_1^p} 
    \\[2mm]
    K_{+2}^{\iota_1^{p+1}} E^{\iota_1^p} 
    = \epsilon^{-p-1} q^{-1}  E^{\iota_1^{p+1}} K_{+2}^{\iota_1^p} 
    \\[2mm]
    K_{-2}^{\iota_1^{p+1}} E^{\iota_1^p} 
    = \epsilon^{-p-1} q E^{\iota_1^{p+1}} K_{-2}^{\iota_1^p} 
  \end{array}
  \qquad\qquad
  \begin{array}{l}
    K_{+1}^{\iota_1^{p+1}} F^{\iota_1^p} 
    = \epsilon^{p+1} q^{-1} F^{\iota_1^{p+1}} K_{+1}^{\iota_1^p} 
    \\[2mm]
    K_{-1}^{\iota_1^{p+1}} F^{\iota_1^p} 
    = \epsilon^{p+1} q F^{\iota_1^{p+1}} K_{-1}^{\iota_1^p} 
    \\[2mm]
    K_{+2}^{\iota_1^{p+1}} F^{\iota_1^p} 
    = \epsilon^{p} q F^{\iota_1^{p+1}} K_{+2}^{\iota_1^p} 
    \\[2mm]
    K_{-2}^{\iota_1^{p+1}} F^{\iota_1^p} 
    = \epsilon^{p} q^{-1} F^{\iota_1^{p+1}}K_{-2}^{\iota_1^p} 
  \end{array}
\end{equation}
\begin{eqnarray}
  \label{eq:EF}
  &&
  \epsilon^{-p } ~ E^{\iota_1^{p+1}} F^{\iota_1^p} 
  - \epsilon^{p+1} ~  F^{\iota_1^{p+1}} E^{\iota_1^p} = 
  (q-q^{-1}) \left( K_{-1}^{\iota_1^{p+1}} K_{+2}^{\iota_1^p} 
    - K_{+1}^{\iota_1^{p+1}} K_{-2}^{\iota_1^p}
  \right) \;,
\end{eqnarray}
We further define
\begin{eqnarray}
  \label{eq:composite1}
  &&
  e^{\iota^p} 
  \equiv  E^{\iota_1^{p-1}} \left(K_{+1}^{\iota_1^{p-1}}\right)^{-1}
  =  \epsilon^{-p+1}q \left(K_{+1}^{\iota_1^{p}}\right)^{-1} E^{\iota_1^{p}} 
  \\
  \label{eq:composite2}
  &&
  f^{\iota^p} 
  \equiv  F^{\iota_1^{p-1}} \left(K_{-1}^{\iota_1^{p-1}}\right)^{-1}
  =  \epsilon^{p}q \left(K_{-1}^{\iota_1^{p}}\right)^{-1} F^{\iota_1^{p}} 
  \\
  \label{eq:composite3}
  &&
  k_{-1}^{\iota^p} 
  \equiv  K_{-1}^{\iota_1^{p-1}} \left(K_{+1}^{\iota_1^{p-1}}\right)^{-1}
  =  \left(K_{+1}^{\iota_1^{p}}\right)^{-1} K_{-1}^{\iota_1^{p}} 
  \\
  \label{eq:composite4}
  &&
  k_{2}^{\iota^p} 
  \equiv  K_{+2}^{\iota_1^{p-1}} \left(K_{+1}^{\iota_1^{p-1}}\right)^{-1}
  =  \epsilon \left(K_{+1}^{\iota_1^{p}}\right)^{-1} K_{+2}^{\iota_1^{p}} 
  \\
  \label{eq:composite5}
  &&
  k_{-2}^{\iota^p} 
  \equiv  K_{-2}^{\iota_1^{p-1}} \left(K_{+1}^{\iota_1^{p-1}}\right)^{-1}
  =  \epsilon \left(K_{+1}^{\iota_1^{p}}\right)^{-1} K_{-2}^{\iota_1^{p}} 
\end{eqnarray}
These operators fulfil the relations
\begin{eqnarray}
  \label{eq:relekfk}
  \begin{array}{l}
    k_{-1}^{\iota^p} e^{\iota^{p'}} = \delta_{p,{p'}}~ q^{-2}~ e^{\iota^p}
    k_{-1}^{\iota^p}
    \\[2mm]
    k_{2}^{\iota^p} e^{\iota^{p'}} = \delta_{p,{p'}}~ q^{-2}~ e^{\iota^p}
    k_{2}^{\iota^p}
    \\[2mm]
    k_{-2}^{\iota^p} e^{\iota^{p'}} = \delta_{p,{p'}}~ e^{\iota^p}
    k_{-2}^{\iota^p}
  \end{array}
  \qquad\qquad
  \begin{array}{l}
    k_{-1}^{\iota^p} f^{\iota^{p'}} = \delta_{p,{p'}}~ q^{2}~ f^{\iota^p}
    k_{-1}^{\iota^p}
    \\[2mm]
    k_{2}^{\iota^p} f^{\iota^{p'}} = \delta_{p,{p'}}~ q^{2}~ f^{\iota^p}
    k_{2}^{\iota^p}
    \\[2mm]
    k_{-2}^{\iota^p} f^{\iota^{p'}} = \delta_{p,{p'}}~ f^{\iota^p}
    k_{-2}^{\iota^p}
  \end{array}
\end{eqnarray}
\begin{equation}
  \label{eq:relef}
  \left[ e^{\iota^p}, f^{\iota^{p'}} \right] = \delta_{p,{p'}}~ (q^2-1)~ 
  \left( k_2^{\iota^p} - k_{-2}^{\iota^p} k_{-1}^{\iota^p} \right)
\end{equation}
and, together with $K_{1}^{\iota^p}$
\begin{eqnarray}
  \label{eq:K1efk}
  && K_1^{\iota^p} e^{\iota^{p'}} = \delta_{p,{p'}}~ 
    \epsilon^{-p+1}~ q~ e^{\iota^{p+1}} K_1^{\iota^p} 
    \\
  && K_1^{\iota^p} f^{\iota^{p'}} = \delta_{p,{p'}}~ 
    \epsilon^{p}~ q^{-1}~ f^{\iota^{p+1}} K_1^{\iota^p}
    \\
  && K_1^{\iota^p} k_{-1}^{\iota^{p'}} = \delta_{p,{p'}}~ 
    k_{-1}^{\iota^{p+1}} K_1^{\iota^p}
    \\
  && K_1^{\iota^p} k_2^{\iota^{p'}} = \delta_{p,{p'}}~ 
    \epsilon~ k_2^{\iota^{p+1}} K_1^{\iota^p}
    \\
  && K_1^{\iota^p} k_{-2}^{\iota^{p'}} = \delta_{p,{p'}}~ 
    \epsilon~ k_{-2}^{\iota^{p+1}} K_1^{\iota^p}
\end{eqnarray}
\begin{proposition}
  The composite operators 
  \begin{eqnarray}
    \label{eq:composl2}
    && \cE \equiv \frac{1}{q-q^{-1}}
    \sum_{p=0}^{P-1} \epsilon^{-\frac12 p(p-3)} q^p
    e^{\iota^p} ,
    \qquad 
    \cF \equiv \frac{1}{q-q^{-1}}
    \sum_{p=0}^{P-1} \epsilon^{\frac12 p(p-1)} q^{-p-1}
    f^{\iota^p} ,
    \\
    &&\cK_{-1} \equiv \sum_{p=0}^{P-1}   k_{-1}^{\iota^p} ,
    \qquad\quad
    \cK_2 \equiv \sum_{p=0}^{P-1} \epsilon^{p}  k_2^{\iota^p}, 
    \qquad\quad
    \cK_{-2} \equiv \sum_{p=0}^{P-1} \epsilon^{p} k_{-2}^{\iota^p} 
  \end{eqnarray}
  satisfy relations isomorphic to those of $\cU_q(gl(2))$, i.e. 
  \begin{eqnarray}
    \label{eq:relEKFK}
    \begin{array}{l}
      \cK_{-1} \cE = q^{-2}~ \cE
      \cK_{-1}
      \\[2mm]
      \cK_{2} \cE = q^{-2}~ \cE
      \cK_{2}
      \\[2mm]
      \cK_{-2} \cE = \cE
      \cK_{-2}
    \end{array}
    \qquad\qquad
    \begin{array}{l}
      \cK_{-1} \cF = q^{2}~ \cF
      \cK_{-1}
      \\[2mm]
      \cK_{2} \cF = q^{2}~ \cF
      \cK_{2}
      \\[2mm]
      \cK_{-2} \cF = \cF
      \cK_{-2}
    \end{array}
  \end{eqnarray}
  \begin{equation}
    \label{eq:relecF}
    \left[ \cE, \cF\right] =  
    \frac{1}{q-q^{-1}}
    \left( \cK_2 - \cK_{-2} \cK_{-1} \right)
  \end{equation}
($\cK_{-2}$ and $\cK_2 \cK_{-1}$ being central.)
\end{proposition}
\textbf{Proof:} By direct check. \finproof
\medskip
\\
The supplementary operator 
\begin{equation}
  \label{eq:ell}
  \ell = \sum_{p=0}^{P-1} ~~\prod_{r=0,..,P-1}^{\longleftarrow} K_1^{(p+r)} 
\end{equation}
(with by convention $K_1^{(a+P)}\equiv K_1^{(a)}$)
is central for odd $P$ and it acts as ``~$i^h$~'' for even
$P$, i.e. it anticommutes with $\cE$, $\cF$ and commutes with
$\cK_{-1}$, $\cK_{\pm 2}$. This operator corresponds for even $P$ to
the existence of another Cartan generator related to a second
deformation parameter $q'=i$, as observed in \cite{ASSS}. Most general
multiparameter deformations were considered in \cite{RD}.

The operators $\cE$, $\cF$ and  $\cK_{-1}$, $\cK_{\pm 2}$ are more
clearly understood when written in a matrix form reflecting the
$\ZZ_P$-grading 
\begin{eqnarray}
  \label{eq:composMatrix}
  \cE = 
  \pmatrix
  {e^{(0)} & 0 & \cdots &  \cr 
    0 & e^{(1)} &  & \vdots \cr
    && \ddots & \cr
    & \cdots && e^{(P-1)}
  }
  \qquad\quad
  \cF = 
  \pmatrix
  {f^{(0)} & 0 & \cdots &  \cr 
    0 & f^{(1)} &  & \vdots \cr
    && \ddots & \cr
    & \cdots && f^{(P-1)}
  }
\end{eqnarray}
and similarly for $\cK_{-1}$, $\cK_{\pm 2}$. 
\\
We finally need to introduce the operators $\bar\cK_1$ and $\cX$ defined by
\begin{eqnarray}
  \label{eq:cK1}
  \bar\cK_1 \equiv \sum_{p=0}^{P-2} K_1^{\iota^p} 
  + (K_1)^{-1} (K_1^\iota)^{-1}...(K_1^{\iota^{P-2}})^{-1} \;,
\end{eqnarray}
and
\begin{equation}
  \label{eq:idV}
  \cX \equiv \sum_{p=0}^{P-1} \epsilon^p ~ \mathrm{Id}\big|_{V_p}
\end{equation}
i.e., in matrix form:
\begin{eqnarray}
  \label{eq:WeylMatrix}
  \bar\cK_1 = 
  \pmatrix
  { 0 & 0 & \cdots &  \prod\limits_{p=0}^{P-2} {K_1^{(p)}}^{-1}\cr 
    K_1^{(0)} & 0 &  & \vdots \cr
    0 & K_1^{(1)} &   & \vdots \cr
    && \ddots & \cr
    & \cdots & K_1^{(P-2)}& 0
  }
  \qquad
  \cX = 
  \pmatrix
  { 1 & 0 & \cdots &  \cr 
    0 & \epsilon &  & \vdots \cr
    && \ddots & \cr
    & \cdots && \epsilon^{P-1}
  }
\end{eqnarray}

\begin{proposition}
  The operators $E^{\iota^p}$,
  $F^{\iota^p}$, $K^{\iota^p}_{\pm 1,2}$ 
  generate an algebra isomorphic to 
  $\cW\otimes \cU_q(gl(2))$, where $\cW$ is a $\epsilon$-Weyl
  algebra, i.e.
  \begin{equation}
    \label{eq:weyl}
    \bar\cK_1 ^P = \cX^P = 1 \qquad \qquad 
    \cX \bar\cK_1 = \epsilon ~\bar\cK_1 \cX 
  \end{equation}
  ($\cU_q(gl(2))$ being understood with the supplementary operator $\ell$
  (\ref{eq:ell})).
\end{proposition}
\textbf{Proof:} First one checks that 
\begin{eqnarray}
  \label{eq:tensor}
  &&
  [\bar\cK_1 , \cE] = 0 \qquad\qquad
  [\bar\cK_1 , \cK_{\pm 2}] = 0 \nn\\
  &&
  [\bar\cK_1 , \cF] = 0 \qquad\qquad
  [\bar\cK_1 , \cK_{-1}] = 0 \;.
\end{eqnarray}
$\cX$ also obviously commutes with $\cE$, $\cF$, $\cK_{\pm 2}$ and
$\cK_{-1}$. 
\\
Then it remains to write all the generators $E^{\iota^p}$,
$F^{\iota^p}$, $K^{\iota^p}_{\pm 1,2}$, in terms of $\cE$, $\cF$,
$\cK_{\pm 2}$, $\ell$,  $\bar\cK_1$ and $\cX$, which can be done using
polynomials of $\cX$ for projections on $V_p$ and then inverting the
relations (\ref{eq:composite1})--(\ref{eq:composite5}).

\finproof

\vspace{-.5cm}

\section{Appendix}
\setcounter{equation}{0}
\begin{eqnarray}
\label{tfd1}
t_{12}&=&\ep \left[b_3^{\pr} - 
\frac{b_3b_6^2a_2^{\pr}+b_3^2a_1b_2^{\pr}-a_6b_2^{\pr}-
\delta_2^{\pr}a_6a_1b_3}{a_1a_6a_3a_5} + \frac{b_1^{\pr}}{a_4}\right.\nn\\
&-& \left.  \frac{b_5^{\pr}b_1b_4a_1-a_5^{\pr}b_1}{a_4a_6a_1} - 
\frac{a_6^{\pr}b_3}{a_3a_5}\right],\\
t_{21}&=&\ep^{-1} \left[b_3^{\pr}+
\frac{b_3b_6^2a_2^{\pr}+b_3^2a_6b_2^{\pr}-a_1b_2^{\pr}-
\delta_2^{\pr}a_6a_1b_3}{a_1a_6a_3a_5}+\frac{b_1^{\pr}}{a_1}\right.\nn\\
&-& \left.\frac{b_5^{\pr}b_1b_4a_6-a_5^{\pr}b_1}{a_2a_6a_1}+
\frac{a_6^{\pr}b_3}{a_3a_5}\right],\\
t_{23}&=&\ep
\left[b_4^{\pr}+\frac{b_6(a_2^{\pr}-b_2^{\pr}b_3a_6)}{a_3a_1a_6} 
-\frac{a_1^{\pr}b_4}{a_2a_4} \right.\nn \\
&-&\left.\frac{b_4(b_1^2a_5^{\pr}+a_6b_4b_5^{\pr})-a_1(b_5^{\pr}+
\delta_5^{\pr}a_6b_4)}{a_1a_6a_2a_4}+\frac{b_6^{\pr}}{a_3} \right],\\
t_{32}&=&\ep^{-1}
\left[b_4^{\pr} - \frac{b_6(a_2^{\pr}+b_2^{\pr}b_3a_1)}{a_5a_1a_6} 
+\frac{a_1^{\pr}b_4}{a_2a_4} \right.\nn \\
&+&\left.\frac{b_4(b_1^2a_5^{\pr}-a_1b_4b_5^{\pr})+a_6(b_5^{\pr}-
\delta_5^{\pr}a_1b_4)}{a_1a_6a_2a_4} + \frac{b_6^{\pr}}{a_5} \right],\\
t_{13}&=&\ep^{-1}\left[-\frac{a_2^{\pr}b_3b_6+b_2^{\pr}b_3a_1+
\delta_2^{\pr}b_3b_6^3}{a_1a_6a_5}+\frac{b_1^{\pr}b_4}{a_2}
\right.\nn\\
&-&\left.\frac{a_5^{\pr}b_4b_1+b_5^{\pr}b_1a_6
+\delta_5^{\pr}b_1^3b_4}{a_2a_1a_6} +
\frac{b_6^{\pr}b_3}{a_5}\right],\\
t_{31}&=&\ep \left[-\frac{a_2^{\pr}b_3b_6-b_2^{\pr}b_3a_6+
\delta_2^{\pr}b_3b_6^3}{a_1a_6a_3}+\frac{b_1^{\pr}b_4}{a_4}
\right.\nn\\
&-&\left.\frac{a_5^{\pr}b_4b_1-b_5^{\pr}b_1a_1
+\delta_5^{\pr}b_1^3b_4}{a_4a_1a_6} -
\frac{b_6^{\pr}b_3}{a_3}\right],\\
f_{12}&=&\ep \left[-\frac{a_1a_6(b_2^{\pr}b_3^2a_1-b_2^{\pr}a_6
-\delta_2^{\pr}a_1a_6b_3)-a_3^2b_6^2(a_2^{\pr}b_3+b_2^{\pr}a_1)}{a_1a_6a_3a_5} 
\right.\nn\\
&+&\left.\frac{b_1^{\pr}}{a_2} - \frac{a_5^{\pr}b_1+b_5^{\pr}a_6b_1b_4+
\delta_5^{\pr}b_1}{a_2a_1a_6} + \frac{(a_6^{\pr}+\delta_6^{\pr})b_3}{a_3a_5}
\right],\\
f_{21}&=&\ep^{-1} \left[-\frac{a_1a_6(b_2^{\pr}b_3^2a_6+b_2^{\pr}a_1
-\delta_2^{\pr}a_1a_6b_3)+a_5^2b_6^2(a_2^{\pr}b_3-b_2^{\pr}a_6)}{a_1a_6a_3a_5} 
\right.\nn\\
&+&\left.\frac{b_1^{\pr}}{a_4}+\frac{a_5^{\pr}b_1-b_5^{\pr}a_1b_1b_4+
\delta_5^{\pr}b_1}{a_4a_1a_6}-\frac{(a_6^{\pr}+\delta_6^{\pr})b_3}{a_3a_5}
\right],\\
f_{13}&=&\ep^{-1}\left[\frac{-b_6(b_2^{\pr}a_6+\delta_2^{\pr}a_1a_6b_3
-a_2^{\pr}b_3)}{a_1a_6a_3}+\frac{b_1^{\pr}b_4}{a_4} \right.\nn\\
&-&\left.\frac{a_1b_1(b_5^{\pr}+\delta_5^{\pr}a_6b_4)-b_1b_4a_5^{\pr}}{a_4a_1a_6}
+\frac{b_6^{\pr}b_3}{a_3}\right],\\
f_{31}&=&\ep\left[\frac{b_6(b_2^{\pr}a_1-\delta_2^{\pr}a_1a_6b_3
+a_2^{\pr}b_3)}{a_1a_6a_5}-\frac{b_1^{\pr}b_4}{a_2} \right.\nn\\
&+&\left.\frac{a_6b_1(b_5^{\pr}-\delta_5^{\pr}a_1b_4)+b_1b_4a_5^{\pr}}{a_2a_1a_6}
-\frac{b_6^{\pr}b_3}{a_5}\right],\\
f_{23}&=&\ep\left[-\frac{b_6(a_2^{\pr}+\delta_2^{\pr}+b_2^{\pr}a_1b_3)}
{a_1a_6a_5}+\frac{(a_1^{\pr}-\delta_1^{\pr})b_4}{a_2a_4} \right.\nn\\
&+&\left.\frac{a_1a_6(b_5^{\pr}a_1+\delta_5^{\pr}a_1a_6b_4-b_5^{\pr}a_6b_4^2)
+a_4^2b_1^2(a_5^{\pr}b_4+b_5^{\pr}a_6)}{a_1a_6a_2a_4}+\frac{b_6^{\pr}}{a_5}
\right],\\
f_{32}&=&\ep^{-1}\left[\frac{b_6(a_2^{\pr}+\delta_2^{\pr}-b_2^{\pr}a_6b_3)}
{a_1a_6a_3} - \frac{(a_1^{\pr}-\delta_1^{\pr})b_4}{a_2a_4} \right.\nn\\
&-&\left.\frac{a_1a_6(-b_5^{\pr}a_6+\delta_5^{\pr}a_1a_6b_4+b_5^{\pr}a_1b_4^2)
+a_2^2b_1^2(a_5^{\pr}b_4-b_5^{\pr}a_1)}{a_1a_6a_2a_4} + \frac{b_6^{\pr}}{a_3}
\right],\\
d_{11}&=&a_3^{\pr}
+\frac{a_2^{\pr}}{a_1a_6}+\frac{a_1^{\pr}+\delta_1^{\pr}b_4^2}{a_2a_4}
+\frac{b_5^{\pr}b_4(a_1-a_6)}{a_1a_6a_2a_4},\\
d_{22}&=&a_2^{\pr}+a_1^{\pr}+a_5^{\pr}+a_6^{\pr}+a_4^{\pr}+a_3^{\pr},\\
d_{33}&=&a_4^{\pr}+\frac{b_2^{\pr}b_3(a_6-a_1)}{a_1a_6a_3a_5}
+\frac{a_5^{\pr}}{a_1a_6}+\frac{a_6^{\pr}+\delta_6^{\pr}b_3^2}{a_3a_5},\\
d_{12}&=&\frac{\delta_5^{\pr}a_1a_6-b_2^{\pr}(a_1-a_1)b_3}{a_1a_6a_3a_5}-
\delta_1^{\pr}+\frac{a_5^{\pr}}{a_1a_6}+\frac{a_6^{\pr}}{a_3a_5}-\delta_3^{\pr},\\
d_{23}&=&\frac{a_2^{\pr}}{a_1a_6}+\frac{a_1^{\pr}}{a_2a_4}-\delta_4^{\pr}
+\frac{b_5^{\pr}b_4(a_6-a_1)+\delta_5^{\pr}a_1a_6}{a_1a_6a_2a_4}-\delta_6^{\pr},\\
d_{13}&=& a_2^{\pr}+a_1^{\pr}+a_5^{\pr}+a_6^{\pr},\\
d_{123}&=&\frac{b_2^{\pr}a_1a_6b_3(a_6-a_1)
-\delta_2^{\pr}-a_2^{\pr}b_6^2}{a_1a_6a_3a_5}-\frac{\delta_1^{\pr}}{a_2a_4}\nn\\
&-&\frac{a_5^{\pr}b_1^2+b_5^{\pr}a_1a_6b_4(a_1-a_6)-\delta_5^{\pr}}{a_1a_6a_2a_4}
-\frac{\delta_6^{\pr}}{a_3a_5}.
\end{eqnarray}

\paragraph{Acknowledgements:}
The authors A.S. and T.S. acknowledge the LAPTH for hospitality,
where this work was carried out. T.S. acknowledge also SCOPE
grant of SNF, Volkswagen foundation, INTAS grant 00561
and
A.S INTAS grant 00-390 for partial financial support.

\vspace{-2mm}

\end{document}